\documentclass[sigconf,nonacm,balance=false]{aamas}

\makeatletter
\gdef\@copyrightpermission{
  \begin{minipage}{0.2\columnwidth}
   \href{https://creativecommons.org/licenses/by/4.0/}{\includegraphics[width=0.90\textwidth]{by}}
  \end{minipage}\hfill
  \begin{minipage}{0.8\columnwidth}
   \href{https://creativecommons.org/licenses/by/4.0/}{This work is licensed under a Creative Commons Attribution International 4.0 License.}
  \end{minipage}
  \vspace{5pt}
}
\makeatother

\setcopyright{none}
\acmConference[Preprint]{Preprint}{22 October 2026}{Online}{}
\copyrightyear{2026}
\acmYear{2026}
\acmDOI{}
\acmISBN{}
\submissionType{Preprint}

\usepackage{booktabs}
\usepackage{amsmath}
\usepackage{xcolor}
\usepackage{balance}
\usepackage{catchfilebetweentags}
\usepackage{etoolbox}

\microtypesetup{protrusion=false}

\newcommand{\Pone}{\textsf{P1a}}
\newcommand{\PER}{\textsc{PER}}

\begin{document}

\title{Beyond Division of Labor: How Roles Shape Feedback and Recovery in Multi-Agent Systems}

\author{Chih-Hsuan Yang}
\authornote{These authors are co-first authors and contributed equally to this work.}
\authornote{Corresponding author: Chih-Hsuan Yang (bellayang@anl.gov).}
\affiliation{%
  \institution{Argonne National Laboratory}
  \country{United States}
}
\email{bellayang@anl.gov}

\author{Xinyang Li}
\authornotemark[1]
\affiliation{%
  \institution{University of Houston}
  \country{United States}
}
\author{Rajeev Thakur}
\affiliation{%
  \institution{Argonne National Laboratory}
  \country{United States}
}
\author{Feng Yan}
\affiliation{%
  \institution{University of Houston}
  \country{United States}
}

\begin{abstract}
Roles in multi-agent large language model (LLM) systems allocate decisions as well as work. In a Planner--Executor--Reviewer team, the Reviewer's approval can determine when the team submits an answer and receives corrective feedback. We study both functions by varying role instructions while holding agents, topology, call budget, and ground-truth evaluator constant, using three open-weight models on Omni-MATH and two on SciBench. Removing roles largely eliminates functional specialization without lowering first-pass accuracy. After a failed first answer, however, submission decisions matter: a role-structured Reviewer often withholds approval, so its team submits less often and receives fewer evaluator verdicts. In an exploratory comparison with feedback available to both teams, the role-free team recovers more often across all five model--benchmark settings. Two further analyses support the importance of feedback opportunities. In a post hoc comparison at equal numbers of verdicts, the pooled recovery difference is small (role-structured minus role-free: $+0.009$, 95\% interval $[-0.014,+0.033]$). In an exploratory intervention on $400$ Omni-MATH problems with gemma-4, keeping the role prompts but forcing submission in every cycle raises recovery from $0.198$ to $0.613$; the paired comparison detects no difference from the role-free team. The takeaway is to evaluate roles through recovery after failure as well as specialization. For systems with reliable external feedback, this requires measuring who may submit, how often the team receives feedback, and which agents can act on it.
\end{abstract}


\begin{teaserfigure}
\noindent\makebox[\textwidth][c]{\small
\href{https://chihhsuan-yang.github.io/AAMAS2027-AgentProtocols/}{Project Website}\quad$\vert$\quad
\href{https://github.com/ChihHsuan-Yang/AAMAS2027-AgentProtocols}{Code}\quad$\vert$\quad
\href{https://huggingface.co/datasets/AgentsSci/AAMAS2027-Roles}{Dataset}
}
\Description{Links to the project website, code, and dataset.}
\end{teaserfigure}

\maketitle

\section{Introduction}
\label{sec:intro}

A Planner decomposes a problem, an Executor carries out the plan, and a
Reviewer critiques the answer: this division of labor is a familiar design
for multi-agent large language model systems~\cite{hong2024metagpt,qian2024chatdev,wu2024autogen}.
Yet a role prompt specifies more than the work an agent should do. In a
Planner--Executor--Reviewer (\PER{}) team, the Reviewer can also decide
whether an answer is submitted for external evaluation. This decision
controls an opportunity to learn from a mistake. Our central finding is that
roles can shape failure recovery through these decision rights even when
their intended division of labor provides no first-pass accuracy gain in
the tested settings.

Role instructions do not by themselves establish that agents specialize.
A model may honor the intended division of labor or solve the entire
problem in every role. We therefore measure both the work agents actually
do and the decisions that determine their access to corrective information.
A final accuracy score alone cannot separate these two functions.

The evaluator in our experiments has access to the reference answer and
returns a verdict and, after a failure, a repair hint. Internal deliberation
has no access to this external signal. A conservative Reviewer saves
evaluations, which can be valuable when they are expensive, but also limits
opportunities to obtain corrective feedback. We ask how this trade-off
affects recovery under a fixed call budget.

To isolate the effect of roles, we fix the agent topology, the actor model
(the language model that plays all three agents), the call budget, and the
ground-truth evaluator. We then vary the organization in two directions.
First, we completely delete every role instruction. Second, we strictly
enforce the division of labor at the interface using typed message formats.
Because feedback typically reaches each agent through its prompt, a role
ablation also rewrites the feedback channel. We therefore carefully record
the exact information each condition receives (\S\ref{sec:method}). In our
main comparison, both teams have equal access to the evaluator's verdict
and hint. While access is matched, the realized number and content of
verdicts differ. This difference is what roles control.

Deleting roles largely dismantles the intended division of labor; the
role-free prompt asks each agent for its current answer. Consistent with
prior role ablations~\cite{dylan2026equal,orogat2026architectural}, first-pass
accuracy does not decline. The difference emerges after a rejected first
answer. With feedback available to both teams, the role-free team recovers
more often across all five model--benchmark settings and achieves higher
final accuracy. Submission logs show that the role-structured Reviewer
frequently withholds approval, including before any feedback exists, and
its team consequently receives fewer verdicts. A post hoc comparison at
equal numbers of verdicts yields a small pooled recovery difference
($+0.009$, 95\% interval $[-0.014,+0.033]$). In an exploratory intervention
on $400$ Omni-MATH problems with gemma-4, forcing submission in each cycle
while preserving the role prompts raises recovery from $0.198$ to $0.613$;
the paired comparison detects no difference from the role-free team.
Together, these results support examining submission policy alongside
specialization. They do not establish that the two organizations use each
verdict equally well.

Strictly enforcing the division of labor provides a complementary test.
It loosens the Reviewer gate and raises recovery, with a statistically significant gain only
for Llama-3.3. However, $8\%$ to $17\%$ of enforced trajectories exhaust the
allowed regenerations without finishing, so end-to-end accuracy does not
improve. Across these comparisons, role fidelity and useful feedback access
are distinct properties: a team can follow its assigned roles while losing
opportunities to correct an answer.
We summarize our contributions as follows:
\begin{itemize}
\item[(1)] We introduce a budget-matched design that varies organization in
  two directions and strictly documents the information each condition
  receives. This includes a feedback-withheld diagnostic that shows what a
  role ablation measures when it also removes the feedback channel.
\item[(2)] We provide empirical evidence across three actors that roles
  successfully induce functional specialization, and across two domains
  that they do not improve first-pass accuracy.
\item[(3)] We provide evidence that the Reviewer's approval governs the
  team's feedback opportunities. We support this claim using submission
  logs, a post hoc verdict-matched comparison, and an exploratory
  intervention that removes the submission gate while keeping role prompts
  intact.
\item[(4)] We outline implications for evaluating multi-agent
  organizations. We recommend reporting recovery rates after a failed first
  attempt and treating decision rights and information pathways as core
  components of the organizational structure (\S\ref{sec:discussion}).
\end{itemize}

\section{Related Work}
\label{sec:related}

\paragraph{Roles in organizational models.} Organization-centered
multi-agent systems give roles a formal existence independent of their
occupants. Moise$^+$ separates structural, functional and \emph{deontic}
specifications~\cite{hubner2002moise}. OperA makes role adoption an explicit
social contract~\cite{dignum2004opera}, and electronic institutions specify
norms outside the agents~\cite{esteva2001formal}, to be regimented
(violation made impossible) or sanctioned~\cite{grossi2007ubilex}. An LLM
role prompt is the degenerate case: functional and deontic clauses in prose,
with nothing to make them bind. Our role-free condition keeps the positions,
the communication graph and the turn order, but deletes these clauses.
Conversely, our enforced condition regiments the functional part at the
interface. In this paper, adaptability means revising an answer across
attempts on the same problem, not reorganizing the
team~\cite{horling2004survey}. Teamwork models make repairing a failed joint
plan a team obligation~\cite{tambe1997flexible}, which a prompt can state
but not secure.

\paragraph{Role specialization and role drift.} Prompted roles are known to
be unreliable. Agents disobey the role specification across annotated
traces~\cite{cemri2025why}, and \emph{role drift} has a training-time
account~\cite{cao2026modules}. Role-label conditioning fails to induce the
intended specialization~\cite{orogat2026architectural}, and agents collapse
into highly similar rationales~\cite{patel2026representational}.
Differentiation can even arise \emph{without} assigned
roles~\cite{kandoussi2026whoami}. Recent enforcement-side work makes role
scope part of the system structure rather than advice in the
prompt~\cite{stettler2026progressive,ge2026roles}. While role drift and the
decoupling of role quality from task performance are
known~\cite{cao2026modules,yang2026precise}, we measure them under a
controlled ablation.

\paragraph{Redundancy and parity.} Dylan et al.~\cite{dylan2026equal} hold
total LLM calls fixed and find an evolved Planner, Executor and Critic team
no better than an evolved single agent. Zhang et
al.~\cite{zhang2026rethinking} find that many reported multi-agent
efficiency gains arise where one agent suffices. This objection is why we
report a single-agent reference (\S\ref{sec:method}). On the same premise,
Chen et al.~\cite{chen2026agentslimming} prune redundant agents.

\paragraph{Feedback, revision and answer lock-in.} Iterative refinement with
self-generated or tool
feedback~\cite{madaan2023selfrefine,shinn2023reflexion,gou2024critic} and
multi-agent debate~\cite{du2024debate,liang2024divergent} are standard
patterns. Models often repeat earlier outputs under exact verifier
feedback~\cite{jiang2026exactfeedback}. Under self-reflection, this failure
is called Degeneration-of-Thought~\cite{liang2024divergent}, and belief
entrenchment is studied directly~\cite{oh2026entrenchment}. Huang et
al.~\cite{huang2024selfcorrect} report that self-correction degrades
accuracy only in the \emph{intrinsic} setting, where the model judges its
own answer, and improves it when the correct label serves as a stopping
rule~\cite[see also][]{stechly2024selfverification,kamoi2024when}. Our loop
is gated by such an oracle (an evaluator that knows the correct answer),
where more attempts buy more coverage, the share of problems solved by at
least one attempt~\cite{brown2024monkeys}. This is the regime in which the
role-structured team's restraint costs it. Closest to our lever, Chen et
al.~\cite{chen2026relabeling} vary only the \emph{role label} on a
byte-identical erroneous claim and move explicit-correction rates by $23$ to
$93$ percentage points. However, they do so in a single agent without budget
matching.

\paragraph{Coordination versus capability.}
Canedo~\cite{canedo2026architecture} frames structured specification as a
\emph{capability equalizer} whose value falls with model strength.
Theoretical~\cite{yu2026bottleneck} and
complementary~\cite{dochkina2026drop} accounts agree. Our enforced condition
reproduces that direction only under a scoring rule that drops the
trajectories that never finish. Under our pre-registered rule (fixed in
writing before the data existed), it reverses for the gemma-4 model and
vanishes for the other actors (\S\ref{subsec:enforce}).

\paragraph{Prior studies of one trace corpus, and critique uptake.} Three
studies report on one multi-agent trace corpus on the Omni-MATH
benchmark~\cite{jiang2026confidence,yang2026protocol,yang2026precise}. The closest study separates
reviewer \emph{precision} from critique
\emph{uptake}~\cite{yang2026precise}. It finds that a pipeline reviewer
flags real errors more precisely than broadcast-style peer discussion does.
Yet its correct critique far less often changes the next candidate, and
uptake improves when the critique is placed in the solver's working context.
We ask an organizational question these studies leave open: what the role
structure changes beyond the division of labor, including who decides when
the team asks for correction. None of them manipulates the role structure,
checks that the specified division of labor is realized, or matches feedback
access across conditions.
Elsewhere, the failure to revise is almost always measured in a single
agent, leaving no organization to ablate.

\section{Definitions and Measurement}
\label{sec:framework}

\subsection{What ``load-bearing'' means here}

An organization is specified by a tuple: agents, an interaction order, a
message-passing graph, a context-visibility policy, and a \emph{role
contract} per agent. A contract asserts that an agent performs some function
(\textsc{must} decompose the problem) or that it does not (\textsc{must not}
state a final answer). Clauses of the second kind make the specification a
\emph{division} of labor. Because a clause is prose handed to a language
model, the \emph{realized} organization is not guaranteed to match it.
Nothing records which clauses were honored.

\textbf{Definition.} A component of an organization is \textbf{load-bearing}
if removing or altering it reproducibly degrades task performance
\emph{under matched call and information budgets}.

The budget clause ensures we test organization rather than scale. Removing
roles alongside agents, calls or context confounds the evaluation. We apply
this definition \emph{per phase}, because a team must first produce an
answer and then repair it if it is wrong. The information clause covers
every channel through which a signal reaches an agent. In a prompted system,
feedback arrives through the same templates that carry the roles.
Consequently, a role ablation can change both the roles and the feedback
channel. A contrast is matched only when it changes one of them
(\S\ref{sec:method}). Matched access does not mean matched use: how many
verdicts a team draws depends on how often it submits.

\subsection{Measuring the realized organization}

\paragraph{Functional collapse (primary).} The first deliberation cycle (one
pass of Planner, Executor and Reviewer) \emph{collapses} if \emph{all three}
agents assert their own final answers. The pipeline then runs
as three solvers in sequence rather than a division of labor. This metric
targets the property the diagram asserts, not textual overlap. We
measure it on the first cycle so it is not confounded with trajectory
length. It is not applicable to the structurally enforced conditions because
their schema forbids an answer field.

\paragraph{Exact output duplication (secondary, a lower bound).} This occurs
when two or more role messages are byte-identical, that is, the same raw
completion character for character. It
bounds collapse from below (supplement) and is never used to compare actors.

\paragraph{Answer authorship, and computation.} We measure whether a Planner
authors a final answer. Structural enforcement leaves the Planner no field
in which to write one, so this measure serves as the manipulation check.
Separately, we track whether a Planner abstains from \emph{computing},
scored as containing no evaluated numeric equality (\textbf{\Pone{}}, the
measure of pre-registered endpoint P1; \S\ref{subsec:prereg}). Because \Pone{} scores the absence of a pattern, it rewards
short messages. Therefore, it cannot reliably measure a division of labor,
and we use it only as a fidelity check (\S\ref{subsec:prereg}).

\paragraph{First pass and recovery.} \emph{First-pass} accuracy grades the
first answer submitted to the evaluator. It need not come from the first
cycle: for the role-structured team, it may follow internal deliberation
without external feedback. If a trajectory never submits an answer before the budget
runs out, it is graded once on its final answer, and that grade is its first
pass. \emph{Recovery} is the fraction of \emph{failed} trajectories, those
whose first graded answer was wrong, that nonetheless finish correct.
Because recovery conditions on a post-treatment variable (whether the first
answer failed, which the intervention can change), recovery contrasts are
descriptive~\cite{montgomery2018conditioning}. We also report the
main contrast on unconditional final accuracy
(\S\ref{subsec:recovery}). The verdict-matched analysis
(\S\ref{subsec:mechanism}) is conditional by construction.

\paragraph{Validation.} \Pone{} was hand-adjudicated against a definition
frozen \emph{before} the adjudication sheet was opened. It clears a pre-set
agreement threshold of $0.80$ (supplement).

\section{Intervention}
\label{sec:method}

\begin{figure}[t]
\centering
\includegraphics[width=\columnwidth]{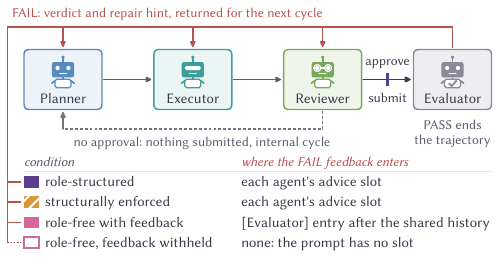}
\Description{A diagram of one deliberation cycle: Planner, Executor and Reviewer in sequence, then an arrow marked approve and submit to the Evaluator. A line from the Evaluator returns the FAIL verdict and repair hint to all three agents in the next cycle; a dashed line from the Reviewer back to the Planner marks a cycle without approval, in which nothing is submitted. Below, a table lists for each of the four conditions who approves submission and through which channel the feedback reaches the agents.}
\caption{\textbf{One deliberation cycle and where the feedback enters.}
A trajectory (one problem's run, at most six cycles) advances one cycle per
pass of Planner, Executor and Reviewer. The third agent's approval (the bar
marks this gate) is the only route to the evaluator, which knows the
reference answer and returns PASS, ending the trajectory, or FAIL with a
repair hint; without approval nothing is submitted and the team deliberates
internally in the next cycle. The key gives where the feedback enters per condition: the advice
slot of each role prompt for the role-structured team (Planner, Executor
and Reviewer prompts) and the structurally enforced one (the same roles
imposed by typed message formats); for the role-free team (one generic
prompt for all three agents, with no slot), an \texttt{[Evaluator]} entry in
the shared history, or nowhere in the feedback-withheld diagnostic. A trajectory that never
submits is graded once when the budget runs out.}
\label{fig:pipeline}
\end{figure}

We vary organization in both directions from one baseline: we remove it, and
we enforce it. All conditions share three agents, one interaction order
(Planner~$\rightarrow$~Executor~$\rightarrow$~Reviewer), the message-passing
graph, context visibility (except under enforcement, below) and the total
call budget. They also use the same verifier and prompt, retry budget,
decoding parameters (temperature $0$, that is, greedy decoding, and
identical \texttt{max\_tokens}), stopping rule (a PASS ends the trajectory)
and grading. Every trajectory in the frozen snapshot (the pre-registered dataset,
\S\ref{sec:experiments}) carries the identical
budget record, so the design tests organization, not scale. Besides the
roles, three things differ. After a FAIL, the evaluator returns a repair
hint, written with the gold answer (the reference solution) in view and
instructed not to reveal it. The role-structured and enforced templates show
the hint to all three agents in the next cycle, each through its role's
advice slot, the template field reserved for the evaluator's feedback
(Figure~\ref{fig:pipeline}). The role-free template has no such slot, so
the role-free team receives the feedback as an entry in its shared history,
and a variant that withholds it serves as a diagnostic. The
role-free footer (the closing instruction of the prompt) also asks every
agent for an answer, and the role-structured Reviewer carries two clauses
the role-free prompt lacks.

\paragraph{Role-structured \PER{} (baseline).} Role prompts are as
ordinarily written. The Reviewer is told to approve only a complete answer
of the right shape. In a cycle that receives a FAIL, it is told to route the
work back to the Planner or Executor rather than approve.

\paragraph{Role-free with feedback (main comparison).} All three agents
receive one byte-identical prompt, ``You are one of three agents
collaboratively solving this problem. Continue the work.'', followed by the
problem, the shared history and the agent's slot number (its position, first
to third). A footer asks for the agent's current best answer in
\texttt{\textbackslash boxed\{\}} and a closing \texttt{[Agree]} (ready to
submit) or \texttt{[Continue]}, the conventions the role-structured Executor
and Reviewer use. Every clause saying who does what is removed, above all
the Planner's \textsc{must not}s on computing and answering. In every prose
condition, the third agent's \texttt{[Agree]} is approval (a message with
neither tag counts as \texttt{[Continue]}), and the latest boxed answer is
submitted. After a FAIL, all three agents see the evaluator's verdict and
repair hint, written by the same evaluator and prompt as for the
role-structured team, as an \texttt{[Evaluator]} entry after the shared
history; like the advice slot, it stays visible for one cycle. The two
teams thus have the same access to feedback; how many verdicts each draws,
and their content, depend on what it submits.

\paragraph{Role-free, feedback withheld (diagnostic).} The same prompt
without the \texttt{[Evaluator]} entry: nothing in this template carries
the evaluator's verdict or hint. It differs from the main role-free
condition only in the feedback and shows what a role ablation measures when
it also removes the feedback channel.

\paragraph{Structurally enforced.} The interface, not the wording, enforces
specialization. The Planner emits JSON (a typed output format) against a
schema with \emph{no answer field}, so a final answer is unrepresentable.
The Executor receives the parsed plan object and the problem, never the
Planner's raw text, and emits a derivation and an answer. The Reviewer emits
a verdict, a target span and a critique. Its \texttt{accept} promotes the
Executor's answer verbatim \emph{by reference}, because it has no field in
which to \emph{author} one. Each function stays with one agent under the
same budget, but the Executor sees less context, and a rejected output can
exhaust the regeneration budget (the number of times a malformed output may
be regenerated). A \emph{context-restored control} repeats this condition
with the shared history restored.

\paragraph{Single-agent reference.} A budget-matched single-agent protocol
on the same problems guards against multi-agent gains that merely mimic a
single capable agent~\cite{zhang2026rethinking}. It comes from a separate,
earlier corpus (\S\ref{sec:experiments}). The comparison cuts both ways, and
the arms come from separate runs (supplement).

\subsection{Instrumentation and pre-registered analyses}
\label{subsec:prereg}

An intervention that never engages cannot explain an outcome, so we measure
engagement. We log every rejected-and-regenerated output with its reason,
count only role-boundary violations in the enforcement rate, and count
answer-bearing calls per condition. One logging gap is disclosed in
\S\ref{subsec:logging}.

All five endpoints (the pre-specified analyses, numbered P1 to P5) were
fixed in writing and committed before any intervention data existed. The full table, with the
outcome for each, is in the supplement. Units are (problem, actor) pairs,
and all conditions run the same problems, so every test is paired, with
bootstrap intervals that resample whole problems~\cite{efron1979bootstrap}.
The pre-registered primary endpoint, P4, asks whether enforcement's effect
on accuracy differs by actor. It is null: no pairwise
difference-in-differences (enforcement's effect in one actor minus its
effect in another) survives Holm's correction for multiple
comparisons~\cite{holm1979simple} (smallest adjusted $p$-value, $p=0.17$).
Before correction, the $95\%$ bootstrap interval for the Llama-3.3 model
against gemma-4, $+0.040$ $[+0.001,+0.080]$, only just excludes zero, and a
separate resample gives $p=0.055$. Only the gpt-oss against gemma-4 contrast had adequate
statistical power for an interaction of plausible size.

The \Pone{} manipulation objective also failed. It required enforcement to
drive Planner computation-abstention to ${\approx}1.0$, but the schema
forbids an \emph{answer}, not a computation. It was re-designated to answer
authorship. The role-structured versus feedback-withheld contrast was
pre-registered as a secondary, exploratory replication of a published null,
and splitting it into phases is post hoc. The role-free condition with
feedback, and with it the main comparison, was added after the freeze (when
the dataset and analysis plan were fixed). Its design and primary analysis
were committed in writing before any of its data existed, but its results
are exploratory, as are those of the forced-submission run
(\S\ref{subsec:mechanism}).
Exclusions were fixed in advance, including the rule that decides
\S\ref{subsec:enforce}: \emph{unparseable trajectories count as incorrect
rather than dropped}.

\subsection{Campaign}
\label{sec:experiments}

\paragraph{Task and problems.} We use the Omni-MATH benchmark of competition
mathematics~\cite{gao2025omnimath}, sampled by a deterministic seeded
manifest (a fixed, reproducible list of problem identifiers) so that every
prefix of the run is a valid random sample. All conditions run the same
problems, so every comparison is paired. A multiple-choice benchmark was
ruled out before launch: on a four-option item, producing a different
candidate succeeds one time in three by chance, which would inflate
recovery.

\paragraph{Actors and scale.} Three actor models are served from one pinned
vLLM server~\cite{kwon2023vllm}: gpt-oss-120b~\cite{openai2025gptoss},
gemma-4-31B~\cite{gemma2026gemma4} and
Llama-3.3-70B~\cite{grattafiori2024llama3}. The pre-registered analysis reads one
frozen snapshot of $11{,}600$ trajectories, $83{,}320$ role messages and
$17{,}486$ gradings on $1{,}600$ problems
sampled uniformly at random from the $4{,}181$ available. The three frozen
conditions run $1{,}600$ problems per cell (one actor in one condition on
one domain) for gemma-4 and gpt-oss and $400$ for Llama-3.3. The
context-restored control covers gemma-4 and gpt-oss at $400$ each. The
evaluator is gpt-oss-120b in every cell, itself one of the actors under
study (\S\ref{sec:limitations}). The single-agent comparator is \emph{not}
part of this campaign; it comes from an earlier, separately collected corpus
on the same benchmark, joined by problem identifier. Power simulated from
that corpus is high for the paired per-actor main effect
(\S\ref{subsec:prereg} for the interaction). An early campaign, stopped when
outputs cut off at the token limit were counted as role-boundary rejections,
was never pooled (supplement).

\paragraph{The role-free condition with feedback.} It was run after the freeze
($4{,}748$ trajectories) on the same server, problems, budget and evaluator,
for all three actors on Omni-MATH and for gemma-4 and gpt-oss on the
SciBench benchmark~\cite{wang2024scibench}. Its data are manifested
separately and never pooled into a frozen cell. The two gemma-4 cells were
run a second time, independently. A re-run (the first of two; supplement) of gemma-4's role-structured
condition on $400$ frozen problems shows no detectable drift in final
accuracy ($0.818$ against $0.825$ frozen, paired $p=0.75$; recovery $0.198$
against $0.146$). Decoding at temperature $0$ is not deterministic on a
shared server. The two role-free conditions are byte-identical until the
first verdict, yet they agree on gemma-4's first verdict for $1{,}496$ of
$1{,}598$ problems, and their first-cycle collapse differs by up to $0.04$
(Table~\ref{tab:collapse}). We therefore treat a cell as one sample and give
the second run, where we have one, in the supplement. Prompts, runner,
analysis scripts and data manifests are in the supplementary material.

\section{Results}
\label{sec:results}

\begin{table}[t]
\centering
\caption{\textbf{Functional collapse on the Omni-MATH benchmark.} Share of
first cycles (one pass of Planner, Executor and Reviewer) in which all three
agents assert their own final answer, per actor (the model playing all
three agents). RS: role-structured (Planner, Executor and Reviewer prompts);
RF$^{-}$ and RF: role-free (one generic prompt for all three agents) with
the evaluator's feedback withheld or shown. \emph{Adjudicated no
division}: the share of a judged sample of first cycles in which a model
judge with a fixed rubric finds no division of labor; the judge did not see
the grade but could infer the condition. The byte-duplication lower bound,
the enforced condition and a sensitivity analysis over answer-marker
definitions are in the supplement.}
\label{tab:collapse}
\small
\begin{tabular}{@{}l ccc cc@{}}
\toprule
 & \multicolumn{3}{c}{functional collapse} & \multicolumn{2}{c}{adjudicated no division} \\
\cmidrule(lr){2-4}\cmidrule(lr){5-6}
actor & RS & RF$^{-}$ & RF & RS & RF$^{-}$ \\
\midrule
gemma-4-31B & $0.000$ & $\mathbf{0.976}$ & $\mathbf{0.974}$ & $0.000$ & $\mathbf{0.857}$ \\
gpt-oss-120b & $0.120$ & $\mathbf{0.846}$ & $\mathbf{0.883}$ & $0.000$ & $\mathbf{1.000}$ \\
Llama-3.3-70B & $0.000$ & $\mathbf{0.872}$ & $\mathbf{0.900}$ & $0.000$ & $\mathbf{1.000}$ \\
\bottomrule
\end{tabular}
\end{table}

\begin{figure}[t]
\centering
\includegraphics[width=\columnwidth]{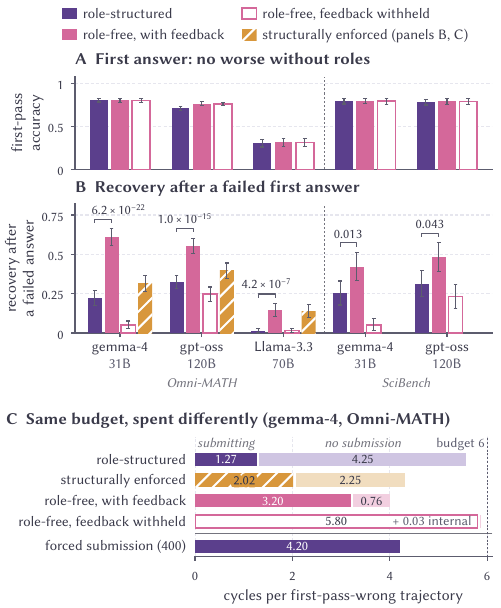}
\Description{Three panels. A: first-pass accuracy per actor and domain, about equal for the role-structured team and both role-free teams except gpt-oss on Omni-MATH, where the role-free teams are higher. B: recovery after a failed first answer; the role-free team with feedback recovers most, the role-structured and enforced teams less, and the role-free team whose feedback is withheld least, tying the role-structured team for Llama-3.3. C: stacked bars of deliberation cycles that end in a submission versus internal cycles without submission for gemma-4 in four conditions, and a fifth bar for the forced-submission run, which submits in every cycle.}
\caption{\textbf{Roles organize the team; the Reviewer's gate limits feedback and recovery.}
Conditions: role-structured (Planner, Executor and Reviewer prompts);
role-free (one generic prompt for all three agents) with the evaluator's
feedback shown, or withheld as a diagnostic (hollow); structurally enforced
(roles imposed by typed message formats; Omni-MATH only). Actors are the
models playing all three agents. \textbf{(A)} First-pass accuracy: whether
a team's first submitted answer is correct. \textbf{(B)} Recovery: the
share of first-pass-wrong trajectories (runs whose first graded answer was
wrong) that finish correct, each condition over its own set; brackets give
the exact McNemar $p$ for role-structured against role-free with feedback,
paired on problems both first answer wrongly. \textbf{(C)} How gemma-4's
first-pass-wrong Omni-MATH trajectories spend the six-cycle budget: cycles
ending in a submission (the third agent approved) or internal cycles
without submission (it declined). The bottom row is the exploratory
forced-submission run on its own $400$-problem subset: role prompts kept,
the answer submitted in every cycle. (A) is over trajectories with a first
verdict; in (B) trajectories that never finish count as unrecovered; error
bars are bootstrap $95\%$ intervals.}
\label{fig:dissociation}
\end{figure}

All statistical contrasts are paired at the problem level, with bootstrap
intervals clustered by problem. Trajectories that never produce a graded
answer count as incorrect, except in the first-pass contrast of
\S\ref{subsec:collapse}. Significance is the exact McNemar
test~\cite{mcnemar1947note} on discordant pairs (problems that exactly one
condition gets right).

\subsection{R1. Roles create specialization, but not a better first
answer}
\label{subsec:collapse}

We delete the role clauses and keep the agents, interaction order, budget
and evaluator; the role-free footer also asks every agent for its current
best answer. Functional collapse, the share of first cycles in which every
agent asserts its own final answer, rises from at most $0.120$ with roles to
at least $0.846$ without them for all three actors on Omni-MATH, the
gemma-4, gpt-oss and Llama-3.3 models (Table~\ref{tab:collapse}). Most of the change comes
from the Planner (per-role rates in the supplement). The Executor answers in
both conditions; the Planner seldom answers with roles (at most $0.183$) and nearly always does
without them, so the division of labor rests on one prohibition, its
\textsc{must not} clause. Because the role-free footer asks for an answer,
the absolute level partly reflects that request; the claim concerns the
\emph{change}. The ordering holds for every actor under all three
answer-marker definitions (the cues that count as asserting an answer;
supplement).

\paragraph{Content duplication and a second opinion.} For the gemma-4 and
Llama-3.3 models, the role-free Planners write full solutions before boxing
an answer, and their messages share far more numbers with the Executor's
than with roles, so the Executor redoes the Planner's work (supplement). A
frontier model judging a sample of first cycles against a frozen rubric
agrees with the collapse measure (Table~\ref{tab:collapse}); the judge was
outcome-blind but not condition-blind (supplement), and the collapse claim
does not rest on it.

\paragraph{First-pass accuracy does not follow.} The collapse costs no
accuracy: on problems graded both with roles and without them (the two
role-free conditions are identical until the first verdict), the observed first-pass accuracy is equal for gemma-4 ($+0.0000$,
$[-0.0156,+0.0150]$), flat for Llama-3.3 and higher without roles for
gpt-oss ($+0.0503$, $[+0.0339,+0.0666]$; Figure~\ref{fig:dissociation}A).
Because the same model plays every agent, solving the problem in the
Planner role does not introduce a different underlying model capability. The gpt-oss model may gain because its
answers vary between derivations (\S\ref{subsec:mechanism}), so extra
independent solutions add chances to be right; we did not test this.

\paragraph{A logging defect.} \label{subsec:logging} Our runner recorded
a response's hidden reasoning trace only when its visible content was
empty, so most of gpt-oss-120b's role-free derivations ($75.7\%$ of
messages) are missing from the record (supplement). Agents pass content
only, in every condition, so the defect is in the record alone and leaves
no point estimate of gpt-oss's duplication. It does not manufacture the
collapse: in first cycles with no affected message ($106$ without roles),
role-free collapse is $0.821$ and role-structured $0.118$.

\subsection{R2. With feedback available to both teams, the role-free team
recovers more}
\label{subsec:recovery}

\paragraph{What a cycle is.} The pipeline runs a fixed grid of six
deliberation cycles, three outer rounds of two inner rounds, each one pass
of Planner, Executor and Reviewer; Figure~\ref{fig:pipeline} shows how a
cycle ends in a \emph{submission} or as an \emph{internal cycle} without
submission.

\begin{table*}[t]
\centering
\caption{\textbf{Role-structured against role-free with feedback, in every cell} (one actor,
the model playing all three agents, on one benchmark; $n$ problems). RS:
role-structured (Planner, Executor and Reviewer prompts); RF: role-free
(one generic prompt for all three agents) with the evaluator's feedback
shown; RF$^{-}$: role-free with it withheld. A trajectory (one problem's
run) is first-pass wrong when its first submitted answer is graded wrong.
\emph{Recovery}: share of first-pass-wrong trajectories that finish
correct, each condition over its own set. \emph{Recovered only by}: on the
$n$ problems both RS and RF first answer wrongly, those recovered by one
condition and not the other, with the exact McNemar $p$ for the two
counts. \emph{Verdict-matched} (post hoc): the same paired recovery with each RF
trajectory cut off after as many evaluator verdicts (PASS or FAIL
decisions) as RS received on that problem, so both get the same number of
verdicts and hints. \emph{Final accuracy}: correct at the end, over all
problems, paired; never-graded trajectories count as incorrect.
\emph{Submissions}: answers the team submits (third-agent approvals) per
first-pass-wrong trajectory, excluding the single end-of-budget grading. \emph{RS never submits}: RS first-pass-wrong trajectories with
no submission in six cycles, graded once at the end.}
\label{tab:main}
\footnotesize
\setlength{\tabcolsep}{2.9pt}
\begin{tabular}{@{}l ccc cccc ccc ccc cc c@{}}
\toprule
 & \multicolumn{3}{c}{recovery} & \multicolumn{4}{c}{recovered only by (paired)}
 & \multicolumn{3}{c}{verdict-matched} & \multicolumn{3}{c}{final accuracy}
 & \multicolumn{2}{c}{submissions} & RS never \\
\cmidrule(lr){2-4}\cmidrule(lr){5-8}\cmidrule(lr){9-11}\cmidrule(lr){12-14}\cmidrule(lr){15-16}
actor, domain ($n$) & RS & RF & RF$^{-}$ & $n$ & RS & RF & $p$ & RS & RF & $p$ & RS & RF & $p$ & RS & RF & submits \\
\midrule
gemma-4, Omni-MATH ($1{,}600$) & $0.222$ & $0.609$ & $0.051$ & $236$ & $4$ & $89$ & $6.2{\times}10^{-22}$ & $0.165$ & $0.153$ & $0.69$ & $0.846$ & $0.921$ & $1.7{\times}10^{-24}$ & $1.27$ & $3.20$ & $84/316$ \\
gpt-oss, Omni-MATH ($1{,}600$) & $0.324$ & $0.552$ & $0.247$ & $319$ & $17$ & $101$ & $1.0{\times}10^{-15}$ & $0.229$ & $0.216$ & $0.72$ & $0.803$ & $0.894$ & $2.7{\times}10^{-28}$ & $2.10$ & $3.53$ & $51/463$ \\
Llama-3.3, Omni-MATH ($400$) & $0.014$ & $0.147$ & $0.015$ & $250$ & $1$ & $26$ & $4.2{\times}10^{-7}$ & $0.012$ & $0.008$ & $1.0$ & $0.315$ & $0.415$ & $1.6{\times}10^{-7}$ & $0.32$ & $2.71$ & $205/276$ \\
\addlinespace
gemma-4, SciBench ($574$) & $0.254$ & $0.419$ & $0.051$ & $100$ & $7$ & $21$ & $0.013$ & $0.230$ & $0.190$ & $0.48$ & $0.847$ & $0.880$ & $0.0054$ & $1.95$ & $3.62$ & $3/118$ \\
gpt-oss, SciBench ($574$) & $0.315$ & $0.483$ & $0.231$ & $93$ & $7$ & $18$ & $0.043$ & $0.215$ & $0.247$ & $0.68$ & $0.852$ & $0.892$ & $1.9{\times}10^{-4}$ & $2.68$ & $3.95$ & $1/124$ \\
\bottomrule
\end{tabular}
\end{table*}

\paragraph{Recovery and final accuracy.} Both teams can obtain the
evaluator's feedback, so this is the main, exploratory comparison
(Table~\ref{tab:main}; \S\ref{subsec:prereg}).
The role-free team recovers more in all five cells. For gemma-4 on
Omni-MATH it recovers $0.609$ of its failed trajectories against $0.222$
(each team over its own; Figure~\ref{fig:dissociation}B), and of the $236$
problems both first answer wrongly, $89$ are recovered only without roles
and $4$ only with them ($p=6.2\times10^{-22}$). Both teams reach the same
first-pass rate and never lose a correct first answer, so the extra
recoveries carry straight into final accuracy ($0.921$ against $0.846$).
They are genuine revisions: nearly every passing answer differs from the
one first rejected, and identical evaluator prompts almost never disagree
(supplement). The other two actors repeat the ordering, and the feedback
leaves their first-pass accuracy essentially unchanged, as expected, since the two
role-free conditions are identical until the first verdict.

\paragraph{Leakage and reproducibility.} A hint written with the gold
answer in view could hand it over. Hints contain it verbatim in at most
$2.3\%$ of failing verdicts in any cell, and dropping the gemma-4
trajectories on Omni-MATH that saw one barely moves either recovery rate
(supplement); we did not audit partial leakage. A second, independent run of the two gemma-4 cells with feedback
is close to the first, though the paired test on the SciBench benchmark
weakens from $p=0.013$ to $0.052$; we report the first run throughout
(supplement).

\subsection{R3. The Reviewer's approval governs feedback opportunities}
\label{subsec:mechanism}

Each submission returns a verdict and, on a FAIL, a hint written with the
gold answer in view, so every submission is another chance to revise. The
two teams differ in how often they seize this opportunity.

\paragraph{A gate on submission.} The role-structured Reviewer dictates when
the team submits. For the gemma-4 model, this team submits $1.27$ times per
failed trajectory, compared to $3.20$ for the role-free team
(Figure~\ref{fig:dissociation}C), and it submits less often in every cell
(Table~\ref{tab:main}). Much of this withholding occurs before any feedback
exists: $53\%$ of the Reviewer's declines happen before the first
submission, and $84$ of the $316$ failed trajectories never submit, so they
are graded once at the end and never see external feedback. The prompt
clause forbidding approval right after a FAIL covers only $19\%$ of the
declines. In later cycles where approval is allowed, the Reviewer still
declines $75\%$ of the time, against $4\%$ for the role-free team's third
agent. On Omni-MATH the clause covers at most $37\%$ of the declines for
any actor, so there the gate reflects the Reviewer's disposition rather
than the letter of its prompt, plausibly because an agent instructed to
critique will search for flaws; on SciBench the clause covers more
(supplement).

\paragraph{Removing the gate.} To test whether submission policy contributes
to the recovery gap, we bypassed the gate while preserving the role prompts.
In a direct, exploratory test, we kept the role-structured prompts byte for byte but
submitted the team's answer in every cycle, overriding the Reviewer. We ran
it once, for gemma-4 on the $400$ Omni-MATH problems of the post-freeze
re-run (supplement). Every cycle the team uses then becomes a submission
(Figure~\ref{fig:dissociation}C, bottom row), and recovery rises from
$0.198$ to $0.613$ (paired, $26$ against $2$, $p=3.0\times10^{-6}$),
with no difference detected from the role-free team's $0.549$ on the same
problems (paired, $7$ against $7$). First-pass accuracy does not move.
This run supports a contribution from submission policy: recovery improves
while the role prompts remain unchanged. The comparison does not establish
equivalence between the two teams.

\paragraph{Matched on verdicts (post hoc).} A descriptive analysis examines
whether the gap persists at an equal number of verdicts. On problems both
teams first answer wrongly, we truncate each role-free trajectory to the
number of verdicts the role-structured team received. No cell then shows a
significant difference (Table~\ref{tab:main}; Llama-3.3's cell has only
five recoveries in total); pooled and stratified by cell, the difference is $+0.009$ in favor
of roles ($95\%$ interval $-0.014$ to $+0.033$). The matching is coarse: a
trajectory cut off after a single FAIL cannot recover, so wherever the
role-structured team received one verdict the pair ties at zero by
construction, as in most of gemma-4's pairs. Restricted to problems with at
least two verdicts, the difference remains not significant; recoding the
scoring defect of \S\ref{sec:limitations} gives $+0.029$ ($+0.004$ to
$+0.054$), a small significant margin for roles (further variants in the
supplement). No variant puts the role-free team significantly ahead. Trajectory truncation cannot establish that the teams use feedback equally
well or quantify how much of the recovery gap submission frequency explains.
Together with the forced-submission run, it supports submission frequency
as one contributor to recovery; feedback use and scoring sensitivity remain
relevant to interpreting the comparison.

\paragraph{The feedback-withheld diagnostic.} This diagnostic removes
external feedback entirely and reveals what a role ablation measures when
it also removes the feedback channel. The role-free team still submits in
almost every cycle (Figure~\ref{fig:dissociation}C), but its agents receive
no signal that the answer failed, so they repeat it: for gemma-4 the team
keeps a single distinct answer in $84.5\%$ of failed trajectories. Adding
the verdict and hint to the identical prompts changes this behavior. The
team explores new answers, and recovery rises from $0.051$ to $0.609$
(paired, $139$ against $2$, $p=7.2\times10^{-39}$). Against this diagnostic
alone, roles would appear load-bearing for repair: strongly for gemma-4
($0.222$ against $0.051$), directionally for gpt-oss ($0.324$ against
$0.247$; $p=0.21$), whose answer moves even without feedback (a single
distinct answer in $20.8\%$ of failed trajectories), and not for Llama-3.3
($0.014$ against $0.015$), whose role-structured team rarely submits.

\subsection{R4. Generality and organizational constraints}
\label{subsec:xdomain}

\paragraph{A second domain.} The SciBench benchmark's complete text-only slice of $574$ free-response
college science problems, frozen separately, tests whether the pattern
belongs to the organization or to competition mathematics. The role-free
prompt is byte-identical across domains; the role-structured and evaluator
prompts adapt only their domain wording (supplement). For both actors run
there, the gemma-4 and gpt-oss models, the ordering returns
(Table~\ref{tab:main}): first-pass accuracy is flat, recovery is highest
with feedback and no roles, and so is final accuracy. The
recovery contrast is weaker, consistent with a looser gate: the
role-structured team almost always submits at least once (all but $3$ and
$1$ of its failed trajectories), so more of the feedback reaches it, and,
as on Omni-MATH, no significant recovery difference is detected after
matching the number of verdicts (answer-lock rates in the supplement).

\paragraph{Structural enforcement.} \label{subsec:enforce} Enforced and role-structured prompts carry the same advice slot, so this
contrast is matched on feedback access. Enforcement replaces the behavioral
request with a constraint: the Planner's final answer becomes
unrepresentable. The manipulation check passes: the Planner asserts a final
answer in $0.000$ of enforced first cycles for all three actors, against at
least $0.889$ without roles. Enforcement does not change what the Planner
\emph{computes}, so the adjudication still finds no functional division in
$0.071$ to $0.188$ of enforced first cycles: an agent can be structurally
stopped from answering without being stopped from doing the work. It does
loosen the gate: the enforced team submits more often for every actor and,
with more hints arriving, unlocks the answer for gemma-4 and Llama-3.3
(supplement).

Recovery rises most where the gate was tightest. Llama-3.3's
role-structured team almost never submits, and under enforcement its
recovery rises about tenfold, from $0.014$ to $0.139$
($p=3.8\times10^{-6}$), close to the role-free team's $0.147$. For gemma-4
and gpt-oss the direction is the same ($p=0.30$ and $0.11$) but not
resolvable against attrition: enforced trajectories that died after an
incorrect first verdict count as unrecovered, as the pre-registration
requires.

Improved recovery does not translate into higher end-to-end accuracy, because
enforcement is brittle. A malformed or boundary-violating output must be
regenerated, and when the regeneration budget runs out the trajectory dies:
$8.0\%$, $9.9\%$ and $17.0\%$ of enforced trajectories never finish,
against at most $2.5\%$ in any other condition, and the weakest actor
breaks the schema most often (supplement). Under the pre-registered rule,
which counts these trajectories as incorrect, enforcement changes accuracy
by $-0.0225$, $-0.0050$ and $+0.0175$ for gemma-4, gpt-oss and Llama-3.3:
one significant negative and two nulls. Dropping them instead gives gains
for all three actors, all significant and ordered by actor weakness,
matching the direction reported for the capability
equalizer~\cite{canedo2026architecture}. That reading is survivorship bias,
since it scores only the trajectories that finished. Restoring the shared
history at fixed enforcement helps gemma-4, mostly through fewer
trajectories dying, and not gpt-oss (supplement).

\section{Discussion}
\label{sec:discussion}

\paragraph{Roles allocate decisions as well as work.} The experiments
separate two functions of role instructions. Removing roles largely
eliminates specialization without reducing first-pass accuracy, whereas
recovery differs when teams use different numbers of feedback opportunities.
The role-structured Reviewer often withholds approval, and its team receives
fewer verdicts, sometimes none. The small pooled difference after matching
verdict counts is descriptive; it does not establish equal revision ability.
The forced-submission intervention provides more direct evidence: for gemma-4
in one exploratory run, recovery improves with the role prompts unchanged.
Together, these observations support treating submission decisions and
feedback access as part of the organization itself.
\paragraph{A design trade-off, not a defect.} A Reviewer that withholds
approval conserves external evaluations, which is valuable when evaluations
are costly, noisy or unavailable. In our setting, with a ground-truth
evaluator and a fixed cycle budget, a strict Reviewer instead limits the
team's opportunities to obtain corrective feedback; teams receiving fewer
verdicts also recover less often. We measure this side of the trade-off, not the full cost
and performance balance. Two design implications follow for similar settings.
First, with reliable external feedback and a fixed call budget, evaluate
submission policy separately from role specialization: keeping role prompts
while relaxing an internal approval gate can improve recovery.
Second, whatever the role structure, route the verifier's feedback to every
agent that can act on it, as the feedback-withheld diagnostic shows
(\S\ref{subsec:mechanism}).

\paragraph{Information audits.} In a prompted multi-agent system, feedback,
tool outputs and verifier verdicts reach the agents through slots in the
same templates that carry the role instructions. Rewriting these templates
during an ablation can therefore silently remove a channel, and no budget
record, call count or accuracy metric reports the absence. A role ablation
thus requires an information audit alongside the budget audit: the runtime
inputs each template consumes must be documented, and a contrast is matched
only when these input sets are identical. Even with matched access, the
realized use can differ, because the organization decides how often the
channels are used.

\paragraph{Decision rights and information pathways.} Prior work shows that
precise reviewing does not guarantee that a critique is
adopted~\cite{yang2026precise}. Similarly, a clean division of labor does
not guarantee that a team recovers from errors. The critical factor is
whether corrective information reaches the team and changes its next
actions. A task-outcome evaluation alone cannot reveal this, because an
aggregate score averages the stages where the organization sits idle with
those where it does its work. We recommend reporting performance
conditioned on a failed first attempt, measuring the realized organization
instead of assuming the specified one, and treating who may submit and
which agents see the feedback as core parts of the organization's
specification.

\section{Limitations}
\label{sec:limitations}

\paragraph{Evaluation paradigm and oracle-gated recovery.} The recovery loop
in our experiments is gated by an oracle. The evaluator triggering retries
sees the ground truth, so every reported recovery is oracle-assisted, resting
on a ground-truth-informed failure signal rather than autonomous
self-correction.
Stopping at the correct answer also hides revisions that would have turned a
correct answer into a wrong one, and an evaluator that can be queried
repeatedly favors the team that queries it most (\S\ref{subsec:mechanism}).

\paragraph{Scope, provenance, and experimental disclosures.} We test one
\PER{} organization, with an evaluator that is itself one of the models
under study. While the ordering returns across a second domain of college
science problems, the findings are limited to text-based reasoning tasks.
The role-free condition with feedback, on which the main comparison rests,
and the forced-submission run are exploratory follow-ups, added after the
freeze and outside the pre-registration. The forced-submission
run covers only one actor, one domain, and a single run, and it removes the
entire gate rather than only the clause forbidding approval after a failure.
Matched access to feedback does not match the realized number or content
of verdicts, and we do not show that any submission policy is cost-optimal.
The condition with feedback also differs in feedback presentation:
role-structured agents receive feedback inside role-specific instructions,
whereas role-free agents view it as a bare \texttt{[Evaluator]} entry.
Finally, the context-restored control is only a \emph{partial} control due
to one-sided attrition (\S\ref{subsec:enforce}), and our functional measure
only detects whether every agent asserted an answer, not whether their
reasoning was identical.

\paragraph{A scoring defect.} The automated runner graded the last
\emph{submitted} answer but recorded the last \emph{held} answer. For the
gemma-4 model on the Omni-MATH benchmark, the two disagree on $17\%$ of
role-structured first-pass-wrong trajectories, and on more in some other
cells (per-cell rates in the supplement). Recoding the affected trajectories
narrows the unadjusted recovery gap and shifts the pooled verdict-matched
difference to $+0.029$ $[+0.004,+0.054]$, which significantly favors the
role-structured team (\S\ref{subsec:mechanism}). The primary analysis keeps
the scoring rule that was fixed before the data were collected.

\section{Conclusion}
\label{sec:conclusion}

Roles govern decision rights and information flow as well as the division
of labor. Across the tested models, removing roles largely eliminated
functional specialization without lowering first-pass accuracy. With
feedback available to both teams, the role-free team recovered more often
in all five model--benchmark settings. Submission logs, a post hoc
verdict-matched analysis, and an exploratory forced-submission intervention
point to the importance of opportunities for external correction. In the
intervention, gemma-4 recovery rose from $0.198$ to $0.613$ while role prompts
stayed unchanged. For systems with reliable external feedback, evaluating
an organization therefore requires measuring who may submit, who receives
the response, and whether the team recovers from a failed first answer.

\begin{acks}
This research used resources of the Argonne Leadership Computing Facility, a
U.S. Department of Energy (DOE) Office of Science user facility at Argonne
National Laboratory (ANL) operated under Contract
No.~DE-AC02-06CH11357. The work was also supported under the same contract by
the DOE Office of Science's Advanced Scientific Computing Research Program and
by Laboratory Directed Research and Development (LDRD) funding from ANL,
provided by the Director, DOE Office of Science.
The work of the University of Houston coauthors was supported in part by the
National Science Foundation under grants CAREER-2305491 and CNS-2504762.
\end{acks}
\bibliographystyle{ACM-Reference-Format}
\bibliography{refs}

\clearpage
\appendix
\setcounter{table}{0}
\setcounter{figure}{0}
\renewcommand{\thetable}{S\arabic{table}}
\renewcommand{\thefigure}{S\arabic{figure}}
\renewcommand{\theHtable}{supp.\arabic{table}}
\renewcommand{\theHfigure}{supp.\arabic{figure}}
\AtBeginEnvironment{table}{\small\setlength{\tabcolsep}{4pt}}
\AtBeginEnvironment{table*}{\small\setlength{\tabcolsep}{4pt}}
\providecommand{\tabnote}[1]{\par\vspace{3pt}\begin{minipage}{\linewidth}\footnotesize\setlength{\parindent}{0pt}#1\par\end{minipage}}
\renewcommand{\topfraction}{0.9}
\renewcommand{\dbltopfraction}{0.9}
\renewcommand{\textfraction}{0.07}
\renewcommand{\floatpagefraction}{0.8}
\renewcommand{\dblfloatpagefraction}{0.8}
\setcounter{topnumber}{3}
\setcounter{dbltopnumber}{3}
\setcounter{totalnumber}{4}
\section*{Supplementary Material}

\section*{What this document is}

The main paper states every claim it makes and is complete without this
file. What it cannot carry in eight pages is the full numeric backing behind
some of its sentences, and that backing is here, in the order of the main
paper's results: the realized organization (R1), recovery (R2), the
submission gate (R3), and the second domain and structural enforcement
(R4), followed by the pre-registered endpoints and the single-agent
comparator. Every table is generated by a script from the manifested data
(the frozen snapshot, or the separately manifested post-freeze runs) with
the same scripts and outputs as the main paper's tables, except the
single-agent comparator, which comes from an earlier, separately collected
corpus. The four figures are generated from the same outputs.

\paragraph{Terms used throughout.} A \emph{trajectory} is one problem's run
through the pipeline, at most six \emph{cycles} of Planner, Executor and
Reviewer. A cycle ends in a \emph{submission} to the evaluator when the
third agent approves the current answer, and in a \emph{decline} otherwise;
each submission returns a \emph{verdict}, PASS or FAIL, and on a FAIL a
repair hint. \emph{First pass} is whether the first submitted answer is
correct; a trajectory that never submits is graded once when its budget
runs out, and that grade is its first pass. \emph{Recovery} is the share of
\emph{first-pass-wrong} trajectories that finish correct. Recovery is
reported on two populations that must not be read across: the
\emph{marginal} population is every first-pass-wrong trajectory of a
condition, on that condition's own denominator; the \emph{paired}
population is only the problems first-pass wrong under both conditions
being compared, on one shared denominator, and it is the population every
test of recovery uses. Tests are exact McNemar tests on the problems that
exactly one condition gets right; intervals are $95\%$ bootstrap intervals
that resample problems; where several tests form a family, $p$ values are
corrected with Holm's step-down procedure. Two \emph{scorings} recur: the
\emph{pre-registered} scoring counts a trajectory that never produced a
graded answer as incorrect, and the \emph{as-graded} scoring drops it. The
conditions are abbreviated RS (role-structured: Planner, Executor and
Reviewer prompts), RF (role-free: one generic prompt for all three agents,
with the evaluator's feedback shown), RF$^{-}$ (role-free with the feedback
withheld) and SE (structurally enforced: the same roles imposed by typed
message formats). The actors are the gemma-4-31B-it, gpt-oss-120b and
Llama-3.3-70B-Instruct models, each playing all three agents.

\section{The realized organization (R1)}
\label{sec:collapse_full}

\begin{figure}[t]
\centering
\includegraphics[width=\columnwidth]{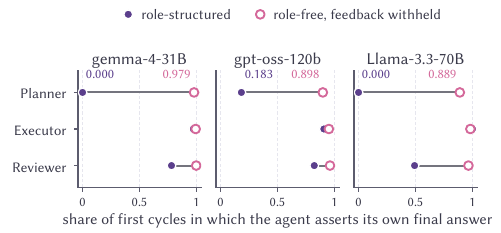}
\Description{Three panels, one per actor. Each shows three horizontal
dumbbells, for the Planner, Executor and Reviewer, from the role-structured
value to the role-free value of the share of first cycles in which that
agent asserts its own final answer. The Planner's dumbbell spans almost the
whole axis in every panel; the Executor's is near one at both ends; the
Reviewer's is short.}
\caption{Which agent changes when the role clauses are deleted. For each
actor, the share of first cycles in which the Planner, Executor and
Reviewer assert their own final answer under role-structured prompts
(filled) and under role-free prompts with the evaluator's feedback withheld
(hollow), on the Omni-MATH benchmark. The Planner moves from almost never
answering to almost always answering; the Executor answers under both; the
Reviewer moves less.}
\label{fig:roles}
\end{figure}

The main paper reports that deleting the role clauses makes every agent
assert its own final answer. Table~\ref{tab:collapse_full} is the full form
of its Table~1, with two further measures and the enforced condition, and
Figure~\ref{fig:roles} shows which agent changes.

\begin{table}[t]
\centering
\caption{Functional collapse on the Omni-MATH benchmark under three measures, with the structurally enforced condition and a sensitivity analysis over answer markers.}
\label{tab:collapse_full}
\begin{tabular}{@{}lccc@{}}
\toprule
 & functional & exact byte & adjudicated \\
actor & collapse & duplication & no division \\
\midrule
\multicolumn{4}{@{}l}{\emph{role-structured (RS)}} \\
\quad gemma-4-31B-it & $0.000$ & $0.000$ & $0.000$ \\
\quad gpt-oss-120b & $0.120$ & $0.000$ & $0.000$ \\
\quad Llama-3.3-70B-Instruct & $0.000$ & $0.000$ & $0.000$ \\
\addlinespace
\multicolumn{4}{@{}l}{\emph{role-free, feedback withheld (RF$^{-}$)}} \\
\quad gemma-4-31B-it & $\mathbf{0.976}$ & $0.000$ & $\mathbf{0.857}$ \\
\quad gpt-oss-120b & $\mathbf{0.846}$ & $0.680^{\dagger}$ & $\mathbf{1.000}$ \\
\quad Llama-3.3-70B-Instruct & $\mathbf{0.872}$ & $0.008$ & $\mathbf{1.000}$ \\
\addlinespace
\multicolumn{4}{@{}l}{\emph{role-free, with feedback (RF)}} \\
\quad gemma-4-31B-it & $\mathbf{0.974}$ & -- & -- \\
\quad gpt-oss-120b & $\mathbf{0.883}$ & -- & -- \\
\quad Llama-3.3-70B-Instruct & $\mathbf{0.900}$ & -- & -- \\
\addlinespace
\multicolumn{4}{@{}l}{\emph{structurally enforced (SE)}} \\
\quad all actors & n/a & $\leq 0.001$ & $0.071$--$0.188$ \\
\bottomrule
\end{tabular}

\tabnote{RS: role-structured prompts; RF$^{-}$ and RF: role-free prompts with the evaluator's feedback withheld or shown (RF was run after the freeze); SE: structural enforcement. Functional collapse: share of first cycles in which all three agents assert their own final answer. Exact byte duplication: share of first cycles in which two or more of the three messages are identical character for character, a lower bound on collapse that is never used to compare actors. Adjudicated no division: share of a judged sample of first cycles in which a model judge with a fixed rubric finds no division of labor; the judge did not see the grade but could infer the condition. Collapse is not applicable under SE, whose Planner schema has no answer field; for RF only functional collapse was measured. \emph{Sensitivity.} Under three answer-marker definitions the
feedback-withheld role-free rate is $0.846$--$0.999$; the
role-structured rate is $0.000$--$0.238$ under the two stricter
markers and $0.328$--$0.892$ under the loosest, which
also counts any ``therefore'', ``hence'' or ``$=$\,number'' (as when a
Planner restates givens), so read the role-structured level as ``low,'' not
a point estimate. The ordering holds in all 9 of 9
actor$\times$marker cells. $^{\dagger}$Not interpretable: $80\%$ of this cell's planner messages are under $500$ visible characters (the main paper's logging defect).}
\end{table}

\paragraph{The Planner is the agent that changes.} By role, the share of
first cycles in which an agent asserts its own final answer moves from RS
to RF$^{-}$ as follows, for gemma-4-31B-it, gpt-oss-120b and
Llama-3.3-70B-Instruct in turn: Planner $0.000$, $0.183$ and $0.000$ to
$0.979$, $0.898$ and $0.889$; Executor $0.973$, $0.907$ and $1.000$ to
$0.993$, $0.951$ and $0.982$; Reviewer $0.781$, $0.823$ and $0.492$ to
$0.998$, $0.961$ and $0.965$. The Executor's job is to produce the answer,
so it answers under both conditions. The Reviewer often states an answer
even with roles, plausibly because approving an answer tends to restate it.
The division of labor therefore rests on the Planner's single prohibition,
and the role-free footer, which asks every agent for its current answer,
lifts it.

\paragraph{Without roles, Planner and Executor compute the same quantities.} The numbers
in a Planner's first-cycle message and in the Executor's overlap more than
Planner and Executor messages from different problems in the same condition
do, by $+0.449$, $+0.680$ and $+0.432$ without roles against $+0.159$,
$+0.167$ and $+0.145$ with them (Jaccard similarity of the two number
sets). Under roles the Planner writes a plan and the Executor computes;
without them both compute the same quantities. The gpt-oss-120b value
without roles is not interpretable, because most of those Planner messages
are only a few characters long (the logging defect below).

\paragraph{A second opinion from a model judge.} A frontier model rated the
first cycles of $171$ sampled trajectories ($42$ RS, $56$ RF$^{-}$ and the
rest SE) against a rubric frozen before the first judgment, asking whether
a division of labor was realized. Its verdicts agree with the
functional-collapse measure (Table~\ref{tab:collapse_full}), and across the
three rubric questions checked for stability its repeated judgments agree
on $0.947$ to $0.959$ of the items. The judge did not see the grade, but it
could infer the condition, because role words were renamed rather than
equalized across conditions; the collapse claim therefore does not rest on
it.

\paragraph{The logging defect, in detail.} A message whose hidden reasoning
was discarded is short but slow, so we mark messages with fewer than $10$
visible characters per second of latency. They are $75.7\%$ of
gpt-oss-120b's role-free messages and $7.7\%$ of its role-structured ones,
and $93.4\%$ and $22.7\%$ of its first cycles contain at least one. The
runner's own fallback flag marks only $5.8\%$ and $2.7\%$ of the same
messages, so the flag undercounts the defect. For the other two actors
at most $0.5\%$ of messages are marked. The defect cannot have produced the
collapse: on first cycles with no marked message, gpt-oss-120b's functional
collapse is $0.821$ ($n=106$) without roles and $0.118$ ($n=1{,}236$) with
roles.

\paragraph{Validating the computation-abstention measure.} \Pone{}, the
pre-registered measure of whether a Planner abstains from computing, was
checked by hand on $72$ Planner messages against a definition frozen before
the adjudication sheet was opened, each judgment with a recorded reason. It
agrees with the hand labels on $0.903$ of messages (Cohen's $\kappa=0.806$;
precision $0.943$, recall $0.868$), clearing the pre-set $0.80$ threshold.
Agreement fell from $0.967$ at $n=30$ as adjudication continued, and we
report the final figure. Its residual errors point in opposite directions
(symbolic results with no digits are missed, restated givens are flagged),
so they do not push the measure one way.

\section{Recovery (R2)}
\label{sec:recovery}

The main paper reports that, with the same access to feedback, the
role-free team recovers more, and that against the feedback-withheld
diagnostic the roles would look load-bearing (the main paper's R3).
Table~\ref{tab:recovery} gives that frozen feedback-withheld comparison on
both populations, so the marginal rates and the paired test
can be checked against each other; the comparison with feedback is the main
paper's Table~2.

\begin{table*}[t]
\centering
\caption{First-pass accuracy and recovery with roles and without them (feedback withheld) on the Omni-MATH benchmark, on both populations.}
\label{tab:recovery}
\begin{tabular}{lcccccc}
\toprule
\multicolumn{7}{@{}l}{\emph{Marginal population: first pass paired; recovery on each condition's own denominator}} \\
\midrule
 & \multicolumn{3}{c}{first pass}
 & \multicolumn{3}{c}{recovery $\mid$ first verdict wrong} \\
\cmidrule(lr){2-4}\cmidrule(lr){5-7}
actor & RS & RF$^{-}$ & difference & RS ($n$) & RF$^{-}$ ($n$) & difference \\
\midrule
gemma-4-31B-it & $0.802$ & $0.802$ & $+0.0000$ {\footnotesize $[{-}0.0156,\,+0.0150]$} & $0.222$ (316) & $0.051$ (316) & ${-}0.1709$ {\footnotesize $[{-}0.2247,\,{-}0.1171]$} \\
gpt-oss-120b & $0.713$ & $0.763$ & $+0.0503$ {\footnotesize $[+0.0339,\,+0.0666]$} & $0.324$ (463) & $0.247$ (380) & ${-}0.0766$ {\footnotesize $[{-}0.1361,\,{-}0.0153]$} \\
Llama-3.3-70B-Instruct & $0.308$ & $0.316$ & $+0.0077$ {\footnotesize $[{-}0.0283,\,+0.0437]$} & $0.014$ (276) & $0.015$ (268) & $+0.0004$ {\footnotesize $[{-}0.0214,\,+0.0192]$} \\
\midrule
\multicolumn{7}{@{}l}{\emph{Paired population: problems first-pass wrong under both conditions; one shared denominator}} \\
\midrule
actor & $n_{\mathrm{both}}$ & RS & RF$^{-}$ & RS only & RF$^{-}$ only & exact $p$ \\
\midrule
gemma-4-31B-it & $237$ & $0.160$ & $0.017$ & $37$ & $3$ & $1.95\times 10^{-8}$ \\
gpt-oss-120b & $327$ & $0.242$ & $0.202$ & $52$ & $39$ & $0.208$ \\
Llama-3.3-70B-Instruct & $245$ & $0.004$ & $0.016$ & $1$ & $4$ & $0.375$ \\
\bottomrule
\end{tabular}

\tabnote{RS: role-structured; RF$^{-}$: role-free with the evaluator's feedback withheld. First pass: whether a trajectory's first submitted answer is correct. Recovery: the share of first-pass-wrong trajectories that finish correct. Upper block: first pass over the problems graded under both conditions; recovery over each condition's own first-pass-wrong trajectories ($n$ in parentheses, not a shared denominator); differences are RF$^{-}$ minus RS, with 95\% bootstrap intervals over problems. Lower block: only the problems first-pass wrong under both conditions ($n_{\mathrm{both}}$), which is the actual test; ``RS only'' and ``RF$^{-}$ only'' count problems recovered under one condition and not the other, with the exact McNemar $p$ for the two counts. The two blocks have different denominators and must not be read across. Recovery conditions on a post-treatment variable and is descriptive rather than causal; no cell recovers before its first evaluator verdict.}
\end{table*}

\paragraph{Where the diagnostic gap comes from.} For gemma-4-31B-it the
accounting closes exactly. The two conditions reach a correct first answer
on the same number of problems ($1{,}284$ of $1{,}600$), and a correct first
answer is never lost because the loop halts on a PASS, so the gap in final
accuracy, $54$ problems in favor of roles ($p=1.6\times10^{-5}$), is exactly
the gap in recoveries, $70$ against $16$. Most of it arises on the $237$
problems both teams first miss, where $38$ are recovered with roles against
$4$ without; the problems that only one team first misses contribute a
further $20$ in the same direction. This accounting needs equal first-pass
counts and does not transfer to the other two actors.

\paragraph{The result reproduces.} The two gemma-4-31B-it cells of the RF
condition were run twice, independently and about thirteen hours apart,
with byte-identical runner and prompt code; every other RF cell was run
once. Table~\ref{tab:bfreplication} gives both runs against the frozen RS
cell. The ordering and the final-accuracy contrasts reproduce in both
domains; the SciBench paired recovery test, on $100$ problems, is $p=0.013$
in the first run and $p=0.052$ in the second, so on that benchmark the
recovery contrast is at the edge of resolution. Each run also repeated the
drift check of the main text on the same $400$ frozen problems: the RS
re-run reaches final accuracy $0.818$ in the first and $0.845$ in the
second, against $0.825$ frozen (paired $p=0.75$ and $p=0.20$). The main
text reports the first run throughout.

\begin{table*}[t]
\centering
\caption{The two gemma-4-31B-it cells of the role-free condition with feedback, run twice, against the frozen role-structured cell.}
\label{tab:bfreplication}
\begin{tabular}{@{}lccccc@{}}
\toprule
 & first pass & recovery & paired recovery & final & paired final \\
\midrule
\multicolumn{6}{@{}l}{\emph{Omni-MATH}} \\
\quad role-structured (frozen) & $0.802$ & $70/316=0.222$ & -- & $0.846$ & -- \\
\quad role-free, with feedback, run 1 & $0.802$ & $193/317=0.609$ & $89$ vs $4$, $p=6.2\times10^{-22}$ & $0.921$ & $137$ vs $17$, $p=1.7\times10^{-24}$ \\
\quad role-free, with feedback, run 2 & $0.811$ & $177/302=0.586$ & $95$ vs $7$, $p=7.8\times10^{-21}$ & $0.922$ & $139$ vs $18$, $p=2.4\times10^{-24}$ \\
\addlinespace
\multicolumn{6}{@{}l}{\emph{SciBench}} \\
\quad role-structured (frozen) & $0.794$ & $30/118=0.254$ & -- & $0.847$ & -- \\
\quad role-free, with feedback, run 1 & $0.796$ & $49/117=0.419$ & $21$ vs $7$, $p=0.013$ & $0.880$ & $31$ vs $12$, $p=0.0054$ \\
\quad role-free, with feedback, run 2 & $0.784$ & $54/124=0.435$ & $19$ vs $8$, $p=0.052$ & $0.878$ & $29$ vs $11$, $p=0.0064$ \\
\addlinespace
\bottomrule
\end{tabular}
\tabnote{Both runs use byte-identical runner and prompt code; the main text reports run 1. First pass: whether a trajectory's first submitted answer is correct. Recovery: first-pass-wrong trajectories that finish correct, over each condition's own first-pass-wrong set (count / denominator). Final: accuracy at the end over all problems, never-graded trajectories counted as incorrect. The paired columns compare each role-free run with the role-structured run on the problems first-pass wrong under both (recovery) or on all problems (final): problems recovered, or finally correct, only without roles versus only with them, with the exact McNemar $p$ for the two counts.}
\end{table*}

\paragraph{Recoveries are revisions, and the hints do not leak the answer.}
Of the $193$ recoveries of the gemma-4-31B-it RF team on Omni-MATH, $192$
pass with an answer different from the one first rejected ($541$ of $547$
across the five RF cells), so re-grading the same answer explains almost
none of them; identical evaluator prompts at temperature $0$ return
different verdicts in only $20$ of $2{,}790$ prompt groups ($0.0072$). The
evaluator writes each hint with the gold answer in view, so a hint could
hand the answer over. Hints contain it verbatim in at most $2.3\%$ of
failing verdicts in any cell, and dropping every gemma-4-31B-it Omni-MATH
trajectory that saw such a hint leaves recovery at $0.599$ for RF and
$0.212$ for RS, against $0.609$ and $0.222$ on all trajectories. We did not
audit partial leakage.

\section{The submission gate (R3)}
\label{sec:gate}

\begin{figure*}[t]
\centering
\includegraphics[width=\textwidth]{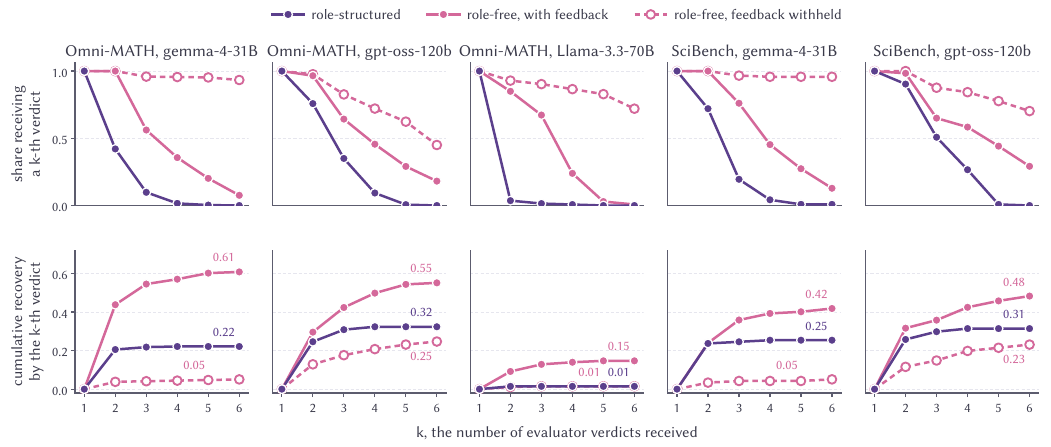}
\Description{Two rows of five panels, one column per cell. Top row: the
share of first-pass-wrong trajectories that receive a k-th verdict, falling
steeply for the role-structured team, more slowly for the role-free team
with feedback, and hardly at all for the role-free team without feedback.
Bottom row: cumulative recovery by the k-th verdict; the role-structured
curve flattens after the second verdict while the role-free curve with
feedback keeps rising.}
\caption{Recovery follows exposure to the evaluator. Top: the share of
first-pass-wrong trajectories that receive a $k$-th evaluator verdict.
Bottom: the cumulative share recovered by the $k$-th verdict, with the
final value labeled. One panel per cell (actor on benchmark):
role-structured (purple), role-free with feedback (rose) and role-free with
the feedback withheld (hollow, dashed). The role-structured recovery curve flattens as fewer trajectories
receive further verdicts; these curves do not isolate how effectively each
team uses a verdict.}
\label{fig:verdicts}
\end{figure*}

\begin{figure*}[t]
\centering
\includegraphics[width=\textwidth]{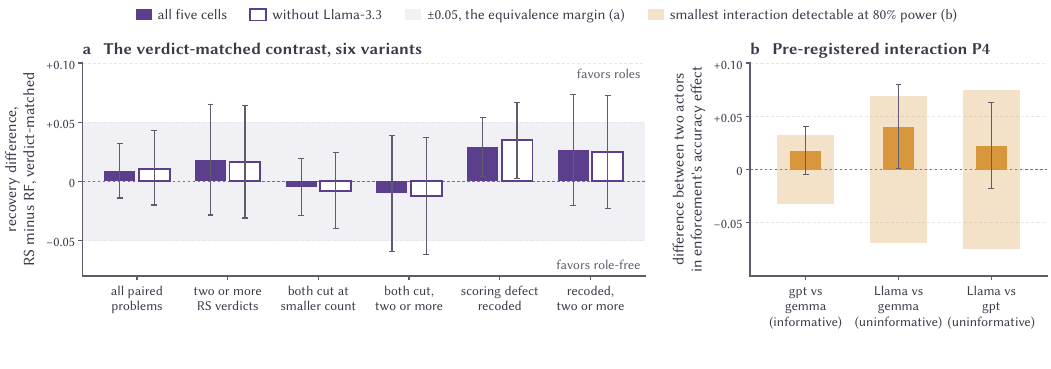}
\Description{Two bar charts with error bars. Left: six pairs of bars,
one pair per definition of the verdict-matched contrast, all small and
with intervals that cross or nearly cross zero, inside a shaded band from
minus 0.05 to plus 0.05. Right: three bars for the pre-registered
interaction, each with an interval, drawn over a wider shaded bar showing
the smallest detectable effect; for two of the three the shaded bar is
wider than the band.}
\caption{Two robustness checks. (a) The verdict-matched recovery
difference, role-structured minus role-free with feedback, pooled over the
five cells (filled) and without Llama-3.3-70B-Instruct (hollow), under the
six definitions of Table~\ref{tab:robust}, with $95\%$ intervals; the band
marks the $\pm0.05$ equivalence margin. (b) The pre-registered interaction
P4 (Section~\ref{sec:prereg}): each difference between two actors in
enforcement's effect on final accuracy, with its $95\%$ interval; the
shaded bar behind it spans the smallest interaction detectable at $80\%$
power. Where that bar is wider than any plausible effect, the null is
uninformative.}
\label{fig:robustness}
\end{figure*}

The main paper's account of what roles do after feedback is that the
role-structured Reviewer gates submission: the RS team asks for fewer
verdicts, and matched on verdicts no significant difference remains. This section
gives the evidence in three steps: where the declines happen
(Table~\ref{tab:gate}), how recovery tracks exposure to the evaluator
(Figure~\ref{fig:verdicts}), and how far the verdict-matched comparison can
be pushed (Table~\ref{tab:robust}, Figure~\ref{fig:robustness}a). The
forced-submission run (Table~\ref{tab:forced}) then removes the gate
directly.

\begin{table*}[t]
\centering
\caption{Where the role-structured Reviewer declines, per cell, and what each team's allocation of the budget buys.}
\label{tab:gate}
\begin{tabular}{@{}l ccc c c cc cc cc cc@{}}
\toprule
 & \multicolumn{3}{c}{RS declining cycles} & clause & RS never
 & \multicolumn{2}{c}{free-cycle approval} & \multicolumn{2}{c}{by 2nd verdict}
 & \multicolumn{2}{c}{per later verdict} & \multicolumn{2}{c}{approvals} \\
\cmidrule(lr){2-4}\cmidrule(lr){7-8}\cmidrule(lr){9-10}\cmidrule(lr){11-12}\cmidrule(lr){13-14}
actor, domain & before & FAIL & free & share & submits & RS & RF & RS & RF & RS & RF & RS & RF \\
\midrule
gemma-4-31B-it, Omni-MATH & $2.24$ & $0.81$ & $1.20$ & $0.19$ & $84/316$ & $0.25$ & $0.96$ & $0.206$ & $0.438$ & $0.41$ & $0.28$ & $1.27$ & $3.20$ \\
gpt-oss-120b, Omni-MATH & $1.33$ & $1.17$ & $0.71$ & $0.37$ & $51/463$ & $0.53$ & $0.68$ & $0.246$ & $0.296$ & $0.27$ & $0.22$ & $2.10$ & $3.53$ \\
Llama-3.3-70B-Instruct, Omni-MATH & $4.49$ & $0.25$ & $0.88$ & $0.04$ & $205/276$ & $0.01$ & $0.82$ & $0.014$ & $0.092$ & $0.25$ & $0.08$ & $0.32$ & $2.71$ \\
gemma-4-31B-it, SciBench & $0.55$ & $1.38$ & $1.48$ & $0.40$ & $3/118$ & $0.35$ & $0.99$ & $0.237$ & $0.239$ & $0.26$ & $0.16$ & $1.95$ & $3.62$ \\
gpt-oss-120b, SciBench & $0.41$ & $1.53$ & $0.60$ & $0.60$ & $1/124$ & $0.67$ & $0.78$ & $0.258$ & $0.317$ & $0.19$ & $0.16$ & $2.68$ & $3.95$ \\
\bottomrule
\end{tabular}
\tabnote{RS: role-structured; RF: role-free with the evaluator's feedback shown. Every column is over first-pass-wrong trajectories. A cycle is one pass of the three agents; it ends in a submission to the evaluator when the third agent approves and in a decline otherwise. RS declining cycles: per trajectory, split into those before the first submission (no feedback yet), those in a cycle that received a FAIL verdict (where the RS clause forbids approval) and the remaining ``free'' cycles; clause share is the FAIL column as a fraction of all three. RS never submits: trajectories with no approval in six cycles, graded once at the end. Free-cycle approval: the share of free cycles ending in approval. By 2nd verdict: the share recovered with at most one round of feedback. Per later verdict: recoveries per verdict after the first, a ratio over different trajectories in the two conditions (the verdict-matched comparison of the main paper's Table~2 is the like-for-like one). Approvals: submissions per trajectory.}
\end{table*}

\paragraph{Where the Reviewer declines.} On Omni-MATH the clause that
forbids approval after a FAIL covers only a minority of the RS Reviewer's
declines, between $4\%$ and $37\%$ depending on the actor, and in the free
cycles, where no clause applies, it approves far less often than the RF
team's third agent (Table~\ref{tab:gate}). The gate is therefore mostly the
Reviewer's disposition rather than the letter of its prompt; gpt-oss-120b's
Reviewer is the most permissive of the three. On SciBench the clause covers
more of the declines, and the RS team almost always submits at least once,
which is why the recovery contrast is weaker there (Section~\ref{sec:xdomain}).

\paragraph{Recovery tracks exposure.} Figure~\ref{fig:verdicts} shows how recovery co-varies with
feedback exposure. In every cell the RS team's share of trajectories still
receiving verdicts falls steeply after the first one, while the RF team's
falls slowly, and the cumulative recovery curves follow the exposure
curves. Per verdict received, neither team is consistently ahead: on the paired
problems, pooled over the five cells, the share whose second verdict is a
PASS is $0.267$ for RS and $0.224$ for RF, and at the third and fourth
verdicts $0.083$ against $0.127$ and $0.105$ against $0.089$. Recoveries
per verdict after the first are higher with roles in every cell
(Table~\ref{tab:gate}), but that ratio compares different trajectories,
since most of the RF team's later verdicts are further attempts on problems
it has already failed more than once; the like-for-like comparison is the
verdict-matched one.

\paragraph{How far the verdict-matched comparison can be pushed.} Matching
on verdicts is coarse. When the RS team received a single verdict, because
it never submitted and was graded at the end or because it submitted once
and stopped, the RF trajectory cut to one verdict cannot recover either,
and the pair ties at zero by construction. These forced ties are most of
the gemma-4-31B-it pairs and nearly all of the Llama-3.3-70B-Instruct pairs
(Table~\ref{tab:robust}), so the pooled interval of the main paper is
narrower than the information behind it. The cut is also one-sided, since
the RS count stops at that team's own passing verdict, which can only favor
it. Table~\ref{tab:robust} and Figure~\ref{fig:robustness}a therefore give
the contrast under six definitions: the problems with at least two RS
verdicts, both trajectories cut at the smaller count, and the scoring
defect recoded, each with and without Llama-3.3-70B-Instruct. No variant
puts the RF team significantly ahead: the two symmetric cuts favor it by
$0.005$ and $0.010$ with intervals that include zero, the informative
problems allow up to $+0.065$ in favor of roles ($+0.073$ with the
recoding), and with the recoding on all problems the RS team leads by
$+0.029$ with an interval that excludes zero. These analyses detect no RF advantage after matching verdict counts,
but they do not establish equal feedback use. The recoded analysis favors
RS, and the estimates depend on the matching and scoring definitions.
Matching equalizes verdict counts rather than verdict content or calls;
the RS team spends more cycles per verdict.

\begin{table*}[t]
\centering
\caption{Robustness of the verdict-matched recovery contrast: forced ties, six definitions of the contrast, and per-verdict pass rates.}
\label{tab:robust}
\begin{tabular}{@{}l ccccc@{}}
\toprule
cell & paired $n$ & forced ties & RS never submitted & problems left & held-answer recodes (RS / RF) \\
\midrule
gemma-4-31B-it, Omni-MATH & $236$ & $141$ & $66$ & $95$ & $12$ / $3$ \\
gpt-oss-120b, Omni-MATH & $319$ & $84$ & $36$ & $235$ & $4$ / $0$ \\
Llama-3.3-70B-Instruct, Omni-MATH & $250$ & $242$ & $187$ & $8$ & $2$ / $1$ \\
gemma-4-31B-it, SciBench & $100$ & $29$ & $2$ & $71$ & $2$ / $0$ \\
gpt-oss-120b, SciBench & $93$ & $11$ & $1$ & $82$ & $0$ / $0$ \\
\bottomrule
\end{tabular}

\vspace{4pt}
\begin{tabular}{@{}l c l c c l@{}}
\toprule
 & \multicolumn{3}{c}{all five cells} & \multicolumn{2}{c}{without Llama-3.3} \\
\cmidrule(lr){2-4}\cmidrule(lr){5-6}
definition & $n$ & difference [95\% interval] & equivalent & $n$ & difference [95\% interval] \\
\midrule
all paired problems (as in the main paper's Table~2) & $998$ & $+0.009$ {\footnotesize $[{-}0.014,\,+0.033]$} & yes & $748$ & $+0.011$ {\footnotesize $[{-}0.020,\,+0.043]$} \\
problems with two or more role-structured verdicts & $491$ & $+0.018$ {\footnotesize $[{-}0.029,\,+0.065]$} & no & $483$ & $+0.017$ {\footnotesize $[{-}0.031,\,+0.064]$} \\
both cut at the smaller verdict count & $998$ & ${-}0.005$ {\footnotesize $[{-}0.029,\,+0.019]$} & yes & $748$ & ${-}0.008$ {\footnotesize $[{-}0.040,\,+0.024]$} \\
both cut, two or more verdicts & $491$ & ${-}0.010$ {\footnotesize $[{-}0.059,\,+0.039]$} & no & $483$ & ${-}0.012$ {\footnotesize $[{-}0.062,\,+0.037]$} \\
scoring defect recoded & $998$ & $+0.029$ {\footnotesize $[+0.004,\,+0.054]$} & no & $748$ & $+0.035$ {\footnotesize $[+0.003,\,+0.067]$} \\
recoded, two or more verdicts & $491$ & $+0.026$ {\footnotesize $[{-}0.020,\,+0.073]$} & no & $483$ & $+0.025$ {\footnotesize $[{-}0.023,\,+0.072]$} \\
\bottomrule
\end{tabular}

\vspace{4pt}
\begin{tabular}{@{}l cc@{}}
\toprule
per-verdict pass rate & role-structured & role-free with feedback \\
\midrule
verdict 2 & $131/491=0.267$ & $212/946=0.224$ \\
verdict 3 & $19/228=0.083$ & $89/700=0.127$ \\
verdict 4 & $8/76=0.105$ & $39/439=0.089$ \\
\bottomrule
\end{tabular}
\tabnote{RS: role-structured; RF: role-free with the evaluator's feedback shown; a cell is one actor on one benchmark. The contrast takes the problems both teams first answer wrongly and cuts the RF trajectory after as many evaluator verdicts as the RS trajectory received on that problem. Where RS received a single verdict, the cut RF trajectory cannot recover either, so the pair ties at zero by construction. Upper block: such forced ties per cell, how many of them never submitted, the problems left, and the trajectories whose held answer matched the key but were graded wrong (the scoring defect of the main paper's Limitations). Middle block: the pooled difference, RS minus RF, stratified by cell with 95\% bootstrap intervals over problems, under six definitions, with and without Llama-3.3-70B-Instruct; equivalent means the 90\% interval lies inside $\pm0.05$. Recoding counts an RS trajectory with a correct held answer as recovered, and an RF one only if it was not cut. Lower block: the share of trajectories whose $k$-th verdict is PASS among those whose earlier verdicts all failed, pooled over the five cells.}
\end{table*}

\paragraph{Across the freeze.} The frozen RS cell and the RF team were run
on opposite sides of the freeze, so we repeat the comparison against the
post-freeze RS re-run of gemma-4-31B-it on $400$ Omni-MATH problems. On the
$65$ problems both first answer wrongly, the RF team leads $22$ to $3$
unmatched ($p=1.6\times10^{-4}$), but the RS team leads $4$ to $1$
verdict-matched ($p=0.38$): the same pattern as in the frozen comparison.

\paragraph{Removing the gate alone.} The forced-submission run tests the
gate directly. It uses the RS prompts and runner unchanged except for one
step: after the Reviewer's turn, the pipeline submits the team's current
answer to the evaluator whatever route the Reviewer chose, and the verdict
and hint reach the agents through the same role-specific slots as in the RS
condition. An offline test in the supplementary material (no model calls)
checks that the prompts are byte-identical to the RS ones up to the first
verdict and that the evaluator is called in every cycle. It was run once,
after the freeze, for gemma-4-31B-it on the $400$ Omni-MATH problems of the
RS re-run, with the same server, budget and evaluator.
Table~\ref{tab:forced} compares it with that re-run, the frozen RS cell and
the RF team, restricted to the same problems. Recovery rises under forced submission, with no detected difference
from RF in the paired comparison and no detected change in first-pass
accuracy. This supports submission policy as a contributor in this setting;
it does not establish equivalence between role prompts and generic prompts.

\begin{table*}[t]
\centering
\caption{Removing the submission gate: gemma-4-31B-it on the same $400$ Omni-MATH problems in every row.}
\label{tab:forced}
\begin{tabular}{@{}l ccccc@{}}
\toprule
condition & first pass & recovery & final accuracy & verdicts & Reviewer approvals \\
\midrule
role-structured, every cycle submitted & $0.768$ & $57/93=0.613$ & $0.910$ & $4.20$ & $0.98$ \\
role-structured, re-run after the freeze & $0.773$ & $18/91=0.198$ & $0.818$ & $1.44$ & $1.13$ \\
role-structured, frozen snapshot & $0.795$ & $12/82=0.146$ & $0.825$ & $1.41$ & $1.15$ \\
role-free with feedback & $0.795$ & $45/82=0.549$ & $0.908$ & $3.34$ & $3.34$ \\
\bottomrule
\end{tabular}

\vspace{4pt}
\begin{tabular}{@{}l ccc cc@{}}
\toprule
 & \multicolumn{3}{c}{recovery (both first wrong)} & \multicolumn{2}{c}{final accuracy} \\
\cmidrule(lr){2-4}\cmidrule(lr){5-6}
pair & $n$ & first / second only & $p$ & first / second only & $p$ \\
\midrule
every cycle submitted vs.\ re-run & $72$ & $26$ / $2$ & $3.0{\times}10^{-6}$ & $42$ / $5$ & $2.5{\times}10^{-8}$ \\
every cycle submitted vs.\ role-free with feedback & $70$ & $7$ / $7$ & $1$ & $8$ / $7$ & $1$ \\
re-run vs.\ role-free with feedback & $65$ & $3$ / $22$ & $1.6{\times}10^{-4}$ & $6$ / $42$ & $1.0{\times}10^{-7}$ \\
\bottomrule
\end{tabular}
\tabnote{Role-structured: Planner, Executor and Reviewer prompts, in the frozen snapshot and in a re-run after the freeze; role-free with feedback: one generic prompt for all three agents, with the evaluator's feedback shown. The forced-submission run (``every cycle submitted'') uses the role-structured prompts byte for byte, but the pipeline submits the team's answer in every cycle whatever the Reviewer decides; its approvals are logged but do not decide submission. First pass: the share of trajectories whose first submitted answer is correct. Recovery: first-pass-wrong trajectories that finish correct, over each row's own first-pass-wrong set (count / denominator). Final accuracy: over all problems, never-graded trajectories counted as incorrect. Verdicts and Reviewer approvals: per first-pass-wrong trajectory, including the first. The lower block pairs two conditions on the problems both first answer wrongly (recovery) and on all problems (final accuracy): problems solved only by the first / only by the second, with the exact McNemar $p$ for the two counts.}
\end{table*}

\section{The second domain (R4)}
\label{sec:xdomain}

Table~\ref{tab:scibench} gives the SciBench feedback-withheld comparison
in full: the frozen RS and RF$^{-}$ cells with their denominators, for the
two actors run there. The main comparison, with feedback, is in the main
paper's Figure~2 and Table~2.

\begin{table*}[t]
\centering
\caption{The SciBench benchmark: first pass and recovery with roles and without them (feedback withheld), two actors, frozen separately.}
\label{tab:scibench}
\begin{tabular}{lccccc@{\hspace{1.4em}}ccccccc}
\toprule
 & \multicolumn{5}{c}{first pass (paired, all problems)}
 & \multicolumn{7}{c}{recovery (paired on first-pass wrong under both)} \\
\cmidrule(lr){2-6}\cmidrule(lr){7-13}
actor & $n$ & RS & RF$^{-}$ & diff. & exact $p$
      & $n_{\mathrm{both}}$ & RS & RF$^{-}$ & diff.
      & RS only & RF$^{-}$ only & exact $p$ \\
\midrule
gemma-4-31B-it & $574$ & $0.794$ & $0.794$ & $+0.0000$ & $1$ & $99$ & $0.212$ & $0.020$ & ${-}0.1919$ & $21$ & $2$ & $6.60\times 10^{-5}$ \\
gpt-oss-120b & $574$ & $0.784$ & $0.789$ & $+0.0052$ & $0.784$ & $96$ & $0.219$ & $0.135$ & ${-}0.0833$ & $17$ & $9$ & $0.169$ \\
\bottomrule
\end{tabular}
\tabnote{The complete 574-problem text-only slice of free-response college science, 2,296 trajectories. RS: role-structured; RF$^{-}$: role-free with the evaluator's feedback withheld; the contrast with feedback shown is in the main paper's Table~2. First pass is paired over all 574 problems, with the exact McNemar $p$ for the problems right under only one condition. Recovery is paired over only the problems first-pass wrong under both conditions ($n_{\mathrm{both}}$); ``RS only'' and ``RF$^{-}$ only'' count problems recovered under one condition and not the other. Differences are RF$^{-}$ minus RS under the pre-registered scoring (never-graded trajectories count as incorrect). gpt-oss-120b's recovery cell is underpowered at its $n_{\mathrm{both}}$; its estimate points the same way as on Omni-MATH. Llama-3.3-70B-Instruct was not run in this domain.}
\end{table*}

\paragraph{What was and was not changed.} The prompts are not byte-identical
to Omni-MATH's, and the prompt files in the supplementary material make the
difference checkable. Of the $16$ prompts the two domains share, $11$ are
identical and $5$ differ: four of the six RS prompts (the Planner, Executor
and Reviewer preambles and the Reviewer's routing rules) and the evaluator's
preamble, which must ask for scientific rather than mathematical
equivalence. Both RF prompts are identical across domains, as are the six SE
prompts (enforcement was not run on SciBench). The asymmetry deserves
stating plainly: the RS prompts, whose team recovers less than the RF team,
are the ones that were rewritten, while the RF prompts ran identical text in
both domains. The rewrite is domain wording (``algebraic manipulation or
numerical calculation'' where Omni-MATH said ``symbolic manipulation or
arithmetic''), and the structural clauses are unchanged; a reader who
suspects the adaptation is doing the work can check the files. Topology,
budget, decoding, retry policy and the evaluator protocol are unchanged, so
this is a replication of the recovery pattern under a matched protocol, not a
rerun of identical prompts.

\paragraph{Why the answer does not lock here.} On Omni-MATH the
gemma-4-31B-it RS team often keeps the same answer across cycles (a single
distinct answer in $28.8\%$ of its failed trajectories); on SciBench it does
not ($7.6\%$, against $8.5\%$ for the RF team). The RS team on SciBench also
almost always submits at least once (Table~\ref{tab:gate}), so it sees more
feedback, and by the second verdict the two teams have recovered equally often. A looser
gate is the simplest common explanation of both observations, though we did
not test it.

\section{Structural enforcement (R4)}
\label{sec:enforcement}

\begin{figure}[t]
\centering
\includegraphics[width=\columnwidth]{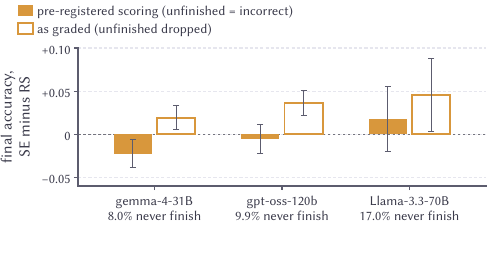}
\Description{A bar chart with error bars. For each of three actors, two
bars of the accuracy difference between the enforced and the
role-structured condition, one under the pre-registered scoring and one as
graded; the as-graded bar is higher than the pre-registered one in every
group. The tick labels give each actor's share of enforced trajectories
that never finish, from 8 percent to 17 percent.}
\caption{Enforcement's effect on final accuracy under the two scorings.
Enforced minus role-structured final accuracy per actor with $95\%$
intervals, under the pre-registered scoring (unfinished trajectories count
as incorrect; filled) and as graded (they are dropped; hollow). The tick
labels give the share of enforced trajectories that never finish, which is
what separates the two bars. The sign flips between the scorings for two
of the three actors, and the as-graded gains are ordered like the
attrition.}
\label{fig:enforcement}
\end{figure}

The main paper reports that enforcement loosens the gate and raises
recovery, but that attrition keeps the gain from reaching end-to-end
accuracy, and that the apparent accuracy benefit under the as-graded scoring
is survivorship. Table~\ref{tab:enforcement} and
Figure~\ref{fig:enforcement} are the evidence for both halves: the
manipulation check, the one-sided attrition, and the same accuracy
difference under both scorings.

\begin{table*}[t]
\centering
\caption{Structural enforcement against the role-structured baseline on the Omni-MATH benchmark: the manipulation check, attrition, and final accuracy under both scorings.}
\label{tab:enforcement}
\begin{tabular}{lccccccc}
\toprule
 & Planner authors & locked answer & recovery & &
 \multicolumn{2}{c}{final accuracy, SE $-$ RS} \\
\cmidrule(lr){6-7}
actor & answer (RF$^{-}$) & (RF$^{-}$ / RS) & (RS) & attrition
      & pre-registered & as graded \\
\midrule
gemma-4-31B-it & $0.000$ (0.986) & $0.119$ (0.845/0.288) & $0.314$ (0.222) & $8.0\%$ & ${-}0.0225$ {\footnotesize [-0.0387,-0.0063]} & $+0.0190$ {\footnotesize [+0.0054,+0.0333]} \\
gpt-oss-120b & $0.000$ (0.916) & $0.092$ (0.208/0.052) & $0.395$ (0.324) & $9.9\%$ & ${-}0.0050$ {\footnotesize [-0.0219,+0.0112]} & $+0.0361$ {\footnotesize [+0.0215,+0.0507]} \\
Llama-3.3-70B-Instruct & $0.000$ (0.958) & $0.072$ (0.806/0.750) & $0.139$ (0.014) & $17.0\%$ & $+0.0175$ {\footnotesize [-0.0200,+0.0550]} & $+0.0453$ {\footnotesize [+0.0030,+0.0876]} \\
\bottomrule
\end{tabular}

\tabnote{SE: structurally enforced; RS: role-structured; RF$^{-}$: role-free with the evaluator's feedback withheld. Planner authors answer: the share of Planner messages that state a final answer, under SE and (in parentheses) under RF$^{-}$; SE drives it to zero in all three actors, so the intervention engaged (the manipulation check). Locked answer: the share of first-pass-wrong trajectories whose answer string never changed, under SE and (in parentheses) RF$^{-}$ / RS. Recovery: the share of first-pass-wrong trajectories that finish correct, under SE and (in parentheses) RS, on the denominator of Table~\ref{tab:recovery}. Attrition: the share of SE trajectories that never finish because a malformed output exhausted its regeneration budget. Final accuracy: the difference over all problems with 95\% bootstrap intervals, under the pre-registered scoring (never-graded trajectories count as incorrect) and as graded (they are dropped); the two columns are different populations and the attrition column is why they disagree. Attrition elsewhere is negligible: $0.0$--$0.8\%$ with roles,
$0.0$--$2.5\%$ without. The as-graded column turns every sign
positive; the pre-registration fixed in advance that unparseable
trajectories count incorrect, and \textbf{the benefit appears only after
conditioning on successful parsing.}}
\end{table*}

\paragraph{Why the columns do not compose.} The enforced recovery and
first-pass rates do not combine into the accuracy column, and cannot:
trajectories that never receive a first verdict have no first pass, and
those that die after one stay in the recovery denominator, so the three
quantities do not share a denominator. The pre-registration fixed in advance
that unparseable trajectories count as incorrect rather than being dropped,
and that rule is what makes the as-graded reading survivorship rather than a
finding.

\paragraph{The gate loosens, and brittleness follows the same order.} Per
failed trajectory, the SE Reviewer approves $2.02$, $2.33$ and $1.84$ times
for gemma-4-31B-it, gpt-oss-120b and Llama-3.3-70B-Instruct, against
$1.27$, $2.10$ and $0.32$ under RS prompts. Enforcement logs $838$,
$1{,}183$ and $395$ role-boundary rejections, that is $0.52$, $0.74$ and
$0.99$ per trajectory, in the same order as the attrition ($8.0\%$, $9.9\%$
and $17.0\%$): the actor that breaks the schema most often is the one that
loses the most trajectories. For Llama-3.3-70B-Instruct, whose RS team
rarely submits, enforced recovery comes close to the RF team's ($0.139$
against $0.147$); on the $219$ problems both first answer wrongly, $9$ are
recovered only under enforcement and $17$ only by the RF team ($p=0.17$).

\paragraph{The context-restored control.} Enforcement also hides the
shared history from the Executor, so part of its brittleness may be lost
context rather than the typed interface. Restoring the shared history
at fixed enforcement (gemma-4-31B-it and gpt-oss-120b, $400$ problems each;
attrition $4.5\%$ and $9.8\%$) changes final accuracy relative to
enforcement by $+0.0450$ $[+0.0125,+0.0775]$ ($p=0.0079$) for gemma-4-31B-it
and by $-0.0125$ $[-0.0475,+0.0200]$ ($p=0.56$) for gpt-oss-120b under the
pre-registered scoring, and by $+0.0196$ $[-0.0028,+0.0419]$ ($p=0.14$) and
$-0.0089$ $[-0.0357,+0.0149]$ ($p=0.65$) as graded. For gemma-4-31B-it the
gain is significant only under the scoring that penalizes attrition, so it
comes mostly from fewer trajectories dying; the control is partial, because
the attrition it removes is one-sided.

\begin{table*}[t]
\centering
\caption{The five pre-registered endpoints and what each returned. P4, the primary, is null before and after Holm correction.}
\label{tab:prereg}
\begin{tabular}{@{}llllll@{}}
\toprule
endpoint & \multicolumn{5}{l}{disposition} \\
\midrule
\textbf{P1} intervention $\rightarrow$ fidelity
 & \multicolumn{5}{l}{\textbf{failed as written; re-designated.} The
   interface forbids an \emph{answer} field, not} \\
\quad \footnotesize(also the manipulation check)
 & \multicolumn{5}{l}{\quad computation. Re-scoped to answer authorship,
   which it drives to $0.000$ in all three} \\
 & \multicolumn{5}{l}{\quad actors (Table~\ref{tab:enforcement}).
   Reported as a second dissociation, not as a pass.} \\
\addlinespace
\textbf{P2} actor $\times$ interv.\ on fidelity
 & \multicolumn{5}{l}{not estimable across the panel: the fidelity
   clauses are condition-specific, so no} \\
 & \multicolumn{5}{l}{\quad common scale exists. Reported as unresolved
   rather than substituted.} \\
\addlinespace
\multicolumn{6}{@{}l}{\textbf{P3} intervention $\rightarrow$ accuracy
  (enforced $-$ role-structured), Holm across three:} \\
 & raw $p$ & Holm $p$ & verdict & & \\
\quad gemma-4-31B-it & $0.0078$ & $0.023$ & reject \\
\quad gpt-oss-120b & $0.60$ & $0.85$ & retain \\
\quad Llama-3.3-70B-Instruct & $0.43$ & $0.85$ & retain \\
\addlinespace
\multicolumn{6}{@{}l}{\textbf{P4} actor $\times$ intervention on
  accuracy (the pre-registered primary), Holm across three:} \\
 & difference-in-diff.\ [CI] & raw $p$ & Holm $p$ & MDE$_{80\%}$ & null is \\
\quad gpt-oss-120b $-$ gemma-4-31B-it & $+0.0175$ {\footnotesize [-0.0050,+0.0406]} & $0.145$ & $0.29$ & $0.0325$ & informative \\
\quad Llama-3.3-70B-Instruct $-$ gemma-4-31B-it & $+0.0400$ {\footnotesize [+0.0006,+0.0800]} & $0.0553$ & $0.166$ & $0.0693$ & uninformative \\
\quad Llama-3.3-70B-Instruct $-$ gpt-oss-120b & $+0.0225$ {\footnotesize [-0.0181,+0.0631]} & $0.286$ & $0.29$ & $0.0748$ & uninformative \\
\addlinespace
\textbf{P5} $\Delta$fidelity vs $\Delta$outcome
 & \multicolumn{5}{l}{association only, pre-declared as non-causal: both
   sides are outcomes of the same} \\
 & \multicolumn{5}{l}{\quad intervention. No $p$-value is reported and
   none is corrected.} \\
\bottomrule
\end{tabular}

\tabnote{The endpoints were fixed in writing before any intervention data existed. The intervention is structural enforcement against the role-structured baseline, per actor. Fidelity: whether the Planner abstains from computing (\Pone{}). Accuracy: final accuracy over all problems, never-graded trajectories counted as incorrect. P3: exact McNemar test per actor, with the raw $p$, the $p$ after Holm's step-down correction across the three actors, and the verdict at $\alpha=0.05$. P4: each difference-in-differences is enforcement's effect in one actor minus its effect in another, with its 95\% bootstrap interval over problems, the raw and Holm-corrected $p$, and MDE$_{80\%}$, the smallest interaction detectable with 80\% power at the realized sample; a null is informative only where the MDE is below a plausible effect. P2 and P5 carry no $p$ value, so the testable contrasts are P3's three and P4's three; Holm across all six changes no verdict. P4's intervals and $p$ values come from two separately seeded bootstrap runs, so an interval can exclude zero while $p$ is just above 0.05; the MDE is computed on problems shared by both actors while each difference-in-differences uses each actor's own problems, so neither can be derived from the other. \textbf{Only one of the three P4 nulls is informative.} The
largest per-actor main effect is small enough that a true interaction is
plausibly ${\leq}0.040$; only the gpt-oss/gemma-4 contrast can see an
effect that size. The two Llama-3.3 contrasts cannot detect an interaction of
any plausible magnitude and are uninformative rather than evidence of
absence.}
\end{table*}

\begin{table}[htb]
\centering
\caption{Recovery by accumulated-context quartile, with roles and without them (feedback withheld), on the Omni-MATH benchmark; directional only.}
\label{tab:context}
\begin{tabular}{@{}lccccc@{}}
\toprule
 & \multicolumn{2}{c}{RS} & \multicolumn{2}{c}{RF$^{-}$} & \\
\cmidrule(lr){2-3}\cmidrule(lr){4-5}
context quartile & recovery & $n$ & recovery & $n$ & difference \\
\midrule
\multicolumn{5}{@{}l}{\emph{gemma-4-31B-it}} \\
\quad Q1 (least context) & $0.662$ & $65$ & $0.151$ & $93$ & $+0.511$ \\
\quad Q2 & $0.280$ & $50$ & $0.009$ & $108$ & $+0.271$ \\
\quad Q3 & $0.047$ & $85$ & $0.014$ & $73$ & $+0.033$ \\
\quad Q4 (most) & $0.078$ & $116$ & $0.000$ & $42$ & $+0.078$ \\
\addlinespace
\multicolumn{5}{@{}l}{\emph{gpt-oss-120b}} \\
\quad Q1 (least context) & $0.923$ & $13$ & $0.258$ & $198$ & $+0.665$ \\
\quad Q2 & $0.656$ & $128$ & $0.293$ & $82$ & $+0.363$ \\
\quad Q3 & $0.262$ & $164$ & $0.191$ & $47$ & $+0.071$ \\
\quad Q4 (most) & $0.070$ & $158$ & $0.189$ & $53$ & ${-}0.119$ \\
\addlinespace
\multicolumn{5}{@{}l}{\emph{Llama-3.3-70B-Instruct}} \\
\quad Q1 (least context) & $0.100$ & $40$ & $0.031$ & $96$ & $+0.069$ \\
\quad Q2 & $0.000$ & $86$ & $0.000$ & $50$ & $+0.000$ \\
\quad Q3 & $0.000$ & $103$ & $0.030$ & $33$ & ${-}0.030$ \\
\quad Q4 (most) & $0.000$ & $47$ & $0.000$ & $89$ & $+0.000$ \\
\addlinespace
\bottomrule
\end{tabular}
\tabnote{RS: role-structured; RF$^{-}$: role-free with the evaluator's feedback withheld. Recovery: the share of first-pass-wrong trajectories that finish correct, with $n$ the number of such trajectories in the quartile; the difference is RS minus RF$^{-}$. Context is the total character count of all agent messages in the trajectory; quartile cut points are computed once per actor over both conditions pooled. Because accumulated context is nearly a function of condition (the RF$^{-}$ team writes less because it stops changing its answer), the strata are not balanced and the comparison cannot separate context from condition.}
\end{table}

\begin{table}[htb]
\centering
\caption{Each three-agent team against a single agent of the same actor under the same call cap, on the Omni-MATH benchmark.}
\label{tab:singleagent}
\begin{tabular}{@{}lccc@{}}
\toprule
organization & team & single agent & difference \\
\midrule
\multicolumn{4}{@{}l}{\emph{gemma-4-31B-it}} \\
\quad role-structured & $0.8462$ & $0.8688$ & ${-}0.0225$ \\
 & \multicolumn{3}{r}{\footnotesize $[{-}0.0400,\,{-}0.0056]$, $p=0.0122$, $n=1600$} \\
\quad role-free & $0.8125$ & $0.8688$ & ${-}0.0563$ \\
 & \multicolumn{3}{r}{\footnotesize $[{-}0.0737,\,{-}0.0394]$, $p=1.22\times 10^{-10}$, $n=1600$} \\
\addlinespace
\multicolumn{4}{@{}l}{\emph{gpt-oss-120b}} \\
\quad role-structured & $0.8031$ & $0.7900$ & $+0.0131$ \\
 & \multicolumn{3}{r}{\footnotesize $[{-}0.0044,\,+0.0300]$, $p=0.152$, $n=1600$} \\
\quad role-free & $0.8194$ & $0.7900$ & $+0.0294$ \\
 & \multicolumn{3}{r}{\footnotesize $[+0.0125,\,+0.0456]$, $p=0.000678$, $n=1600$} \\
\addlinespace
\midrule
\multicolumn{4}{@{}l}{\emph{Realized call budget}} \\
\midrule
actor & protocol & evaluator & actor \\
\midrule
gemma-4-31B-it & single agent & $1.66$ & $3.16$ \\
gemma-4-31B-it & role-structured team & $1.11$ & $6.76$ \\
gemma-4-31B-it & role-free team & $1.95$ & $5.87$ \\
gpt-oss-120b & single agent & $1.69$ & $3.40$ \\
gpt-oss-120b & role-structured team & $1.35$ & $7.54$ \\
gpt-oss-120b & role-free team & $1.85$ & $6.25$ \\
\bottomrule
\end{tabular}
\tabnote{Final accuracy over all problems ($n$) of the role-structured (RS) and role-free, feedback-withheld (RF$^{-}$) teams against a single agent of the same actor solving the same problems under the same cap on model calls; the difference is team minus single agent, with its 95\% bootstrap interval over problems and the exact McNemar $p$. The single-agent trajectories come from an earlier, separately collected trace corpus, joined by problem, and its two actor arms were produced in separate runs: each direction within an actor is a sound guard, the gap between actors is not. Realized call budget: calls per problem to the evaluator and to the actor.}
\end{table}

\paragraph{An early campaign stopped for a truncation confound.} The main
run was gated on a smoke test requiring non-zero boundary rejections with
reasons logged, answer-bearing calls equal across conditions, and a pinned
backend. An early campaign was stopped after $121$ trajectories when $74\%$
of enforcement rejections proved to be the model cut off at the token limit
mid-JSON rather than violating a role boundary. Truncation was not
condition-specific but its cost was: a prose condition tolerates a truncated
message where the enforced one rejects unparseable JSON, so left alone it
would have produced a convincing ``enforcement lowers accuracy'' artifact.
The main run uses twice that token limit, and the early rows were never
pooled.

\paragraph{Accumulated context as a competing explanation.} The RS team's
recovery advantage over the RF$^{-}$ team could in principle come from the
longer history it accumulates over its internal cycles rather than from the feedback
it sees. Table~\ref{tab:context} stratifies recovery by accumulated context.
The comparison is directional only: quartile cut points are computed over
both conditions pooled, and accumulated context is nearly a function of
condition, because the RF$^{-}$ team stops changing its answer, so the strata are not
balanced and context cannot be separated from condition. It is reported
because it does not behave uniformly. The gap survives in the lower
quartiles, but in gpt-oss-120b's top quartile the ordering reverses, as it
does on a handful of events in Llama-3.3-70B-Instruct's third, so more
context is not simply more recovery.

\section{The pre-registered endpoints in full}
\label{sec:prereg}

The main paper reports that the pre-registered primary endpoint P4, whether
enforcement's effect on accuracy differs by actor, is null, and that the
\Pone{} manipulation objective failed. Table~\ref{tab:prereg} is the
complete endpoint table: all five endpoints as written before any
intervention data existed, what each one returned, and the Holm-corrected
verdict.

Two things in it should be read rather than skimmed. First, only one of the
three P4 contrasts is an informative null (Figure~\ref{fig:robustness}b):
the largest per-actor main effect is small enough that a true interaction
is plausibly at most $0.040$, only the gpt-oss-120b against gemma-4-31B-it
contrast can detect an effect of that size, and the two contrasts involving
Llama-3.3-70B-Instruct cannot detect an interaction of any plausible
magnitude at their sample size. They are uninformative rather than evidence
of absence. Second, the contrast pre-registered as secondary and exploratory was RS
against RF$^{-}$; the contrast this paper is organized around, RS against
RF, and the forced-submission run are outside the pre-registration; the RF
condition's design and primary analysis were committed in writing before
its data existed, but its results are exploratory, and the main paper
labels them so.

\section{The single-agent comparator}
\label{sec:singleagent}

Table~\ref{tab:singleagent} is the single-agent reference under the same
call cap. The main paper uses it as a guard against reading ``collapse costs
nothing'' as ``the system was one agent all along'', and reports both
directions. The result cuts both ways: a collapsed gpt-oss-120b team beats
its own single agent, so three agents emitting one answer are not
functionally one agent, while gemma-4-31B-it's team is significantly worse
than its own single agent in both conditions. Neither ``multi-agent systems
are just single agents'' nor ``more agents help'' survives it. The
comparator comes from an earlier, separately collected corpus whose two
actor arms were produced in separate runs, so each direction within an
actor is a sound guard and the gap between actors is not.

\section{AI use statement}

In this work, we used generative AI tools for three purposes: to edit the
manuscript for readability and polish its wording; to search for and identify
relevant literature and to format references; and to assist with data
processing and analysis, including drafting and refining the code for these
tasks and for the figures. All AI-generated text and code were reviewed by the
authors, and LLM-generated code was verified and tested for correctness by the
authors. We take responsibility for the final content of this work, including
text, claims or artifacts produced with the aid of generative AI.

\end{document}